\documentclass[manuscript]{aastex631}
\usepackage{amsmath,amssymb,amsfonts}
\usepackage{hyperref}
\usepackage{multirow}          
\usepackage{enumitem}          
\usepackage{url}               
\usepackage{underscore}        
\usepackage{xcolor}

\makeatletter
\let\frontmatter@title@above\@empty
\makeatother
\begin{document}
\title{Developing Machine Learning Models of Subgrid Turbulent Transport for Quiet Sun 3D Radiative Hydrodynamic Simulations}



\correspondingauthor{Rimsha Hameed Syeda}
\email{rsyeda2@student.gsu.edu}

\author[0009-0007-6965-9152]{Rimsha Hameed Syeda}
\affiliation{Computer Science, Georgia State University, Atlanta, GA, USA}
\author[0000-0002-1837-8365]{Dustin Kempton}
\affiliation{Computer Science, Georgia State University, Atlanta, GA, USA}  
\author[0000-0002-4001-1295]{Viacheslav Sadykov}
\affiliation{Physics and Astronomy, Georgia State University, Atlanta, GA, USA}
\author[0000-0003-4144-2270]{Irina Kitiashvili}
\affiliation{Computational Physics Branch, NASA Ames Research Center, Moffett Field, CA, USA}
\author[0000-0001-9598-8207]{Rafal Angryk}
\affiliation{Computer Science, Georgia State University, Atlanta, GA, USA}





\begin{abstract}

Numerical modeling of solar plasma dynamics is affected by the resolution of the computational grid. This often requires the estimation of subgrid processes related to the small-scale flow turbulence, as these processes play a critical role in momentum transport and energy dissipation. In this work, we investigate the use of deep learning techniques as surrogate models for subgrid turbulent transport in realistic hydrodynamic simulations of the quiet Sun. We describe the development of a 3D Convolutional Neural Network (CNN) to capture spatial dependencies in 3D velocity fields, leveraging different activation functions, as well as different architectural designs. We specifically focus on the prediction of Reynolds stress tensor components. The resultant model integrates velocity vector components and scalar features, such as plasma density, to enhance prediction accuracy. We compare the 3DCNN model to other types of models, such as a Multilayer Perceptron (MLP) and physics-based Gradient and Smagorinsky models, and show that the final model design reconstructs the Reynolds stress tensor components more accurately. Specifically, a 3DCNN model achieves an average improvement of $\sim$31\% on diagonal components and $\sim$8\% on the off-diagonal components of the stress tensor. Additionally, we show that applying a logarithmic data transformation of the target stress tensor components, to handle heavily skewed data, improves model performance. Results demonstrate the potential of deep learning, particularly CNNs, to approximate Reynolds stress tensor components for the upper solar convection zone and lower atmosphere, making them a viable candidate for modeling subgrid processes and a promising alternative to traditional turbulence models.
\end{abstract}



\section{Introduction}
Observations of the Sun reveal a complex multiscale interaction of turbulent radiating plasma in the presence of magnetic fields. Understanding the energy transport between a wide range of scales (from centimeters to thousands of kilometers) is critical for uncovering a variety of observed phenomena (e.g., dynamics of granules and the formation of active regions) and understanding the origin of solar activity. High-resolution observations of the Sun by the Daniel K Inouye Solar Telescope (DKIST) and the Goode Solar Telescope (GST) permit one to probe the solar surface at the spatial scales of $\sim20-50$\, km \citep{Rimmele2020SoPh, Cao2010AN....331..636C}. However, even these high-resolution observations can neither provide sufficient detail on the dynamics of the Sun on the surface, nor can they directly reveal the complex behavior of plasma in the subphotospheric region.

By using state-of-the-art computational capabilities to generate 3D radiative MHD models based on the first physical principles, researchers can generate synthetic observations corresponding to a particular instrument or telescope with more detail, as well as investigate the subphotospheric regions hidden from direct observation. This `ab initio' modeling approach is utilized in several codes, such as StellarBox \citep{Wray2015_stellarbox, Wray2020}, Bifrost \citep{Gudiksen2011A&A...531A.154G}, MURAM \citep{Vogler2005A&A...429..335V}, MANCHA3D \citep{Navarro2022A&A...663A..96N}, etc. In these codes, the effects of plasma compressibility, radiation, and magnetic fields are often taken into account together with the realistic equation of state, abundances, and the structure of the solar interiors and atmosphere. These approaches are capable of reproducing various observational phenomena, as discussed in, e.g., \citep{Kitiashvili2011,Kitiashvili2014,Rempel2018,Bakke2023}. However, when considering a wide range of scales, no model captures the Sun's turbulent dynamics all the way down to the smallest scales. In order to mitigate this issue, Large Eddy Simulations (LES) resolve all essential scales explicitly \citep{Kitiashvili2013}, and utilize a subgrid-scale (SGS) turbulence model to mitigate the impact of energy and momentum transport on subgrid scales. For example, in the StellarBox code that provides the simulation data for our investigation, a compressible version of the Smagorinsky SGS model and its dynamic version are utilized~\citep{Smagorinsky1963_viscosity, Germano1991_dynamicSmagorinsky}. Therefore, the SGS models aim to connect unresolved turbulence to the larger scales that are explicitly simulated, ensuring that large-scale simulations remain computationally feasible without sacrificing essential physical accuracy.


Subgrid turbulence generally refers to the turbulent motions and related interactions (that result in energy dissipation and momentum transport) that occur at scales smaller than the grid resolution of a simulation. Simulations typically maintain a trade-off between computational feasibility (related to the characteristic scales of the modeled phenomena and the number of computational grid cells in each direction) and the need to resolve the high-resolution dynamics explicitly, leaving a treatment of subgrid processes to SGS models. In recent years, machine learning (ML) techniques, in particular, deep learning models, have shown promise in improving SGS models by learning directly from high-fidelity high-resolution simulation data or observations \citep{Panda2021arXiv211107043P, Karpov2022}. Given a key challenge in simulating solar dynamics is balancing the need to accurately capture the physics of the radiating turbulent plasma (which requires high resolution and computational cost) and the feasibility of a computational setup, the development of deep learning-driven SGS models is an important direction to investigate.



This work closely follows the experimental design for data-driven SGS models in \citep{Karpov2022}. In that work, the authors utilized high-resolution simulations of supernova explosions and demonstrated the ability to run lower-resolution simulations by utilizing subgrid estimates of the higher-resolution properties. Their results show that the lower-resolution simulation domain behaves similarly to high-resolution simulations, which provides the critical validation of the subgrid modeling approach. In this investigation, we adopt their approach to simulations of the quiet Sun, utilizing solar radiative hydrodynamic simulations from \citep{Kitiashvili2015} obtained with StellarBox simulation code \citep{Wray2015_stellarbox, Wray2020}.

The primary goal of this work is to investigate data-driven alternatives to physics-motivated turbulence models using machine learning (ML), testing multiple ML model designs to find the most appropriate one for modeling subgrid Reynolds stress tensor components. This type of surrogate modeling, where a simpler model is trained to approximate the behavior of a more complex system, has become an increasingly popular approach in various fields of computational science. Surrogate models can be particularly valuable for reducing the computational cost of these simulations, enabling more extensive parameter studies. In this paper, we aim to explore the potential of deep learning models (especially, 3D Convolutional Neural Networks, 3DCNNs) to serve as a basis for the SGS model for the 3D radiative hydrodynamic simulations of the solar upper convection zone and lower atmosphere.

We begin by providing some background information about the StellarBox simulation and the typical subgrid-scale turbulence models used in Section~\ref{section:Background}. Then we discuss the methodology employed in this work in Section~\ref{section:Methodology}, where we discuss the preparation and preprocessing of the simulation data(Subsection~\ref{section:data_prep}) and the deep learning model designs (Subsection~\ref{section:models}). The experimental settings of models are described in detail in Section~\ref{section:settings} (with hyperparameter tuning of MLP in Subsection~\ref{section:mlp_tuning} and CNN in Subsection~\ref{section:cnn_tuning}). The detailed discussion of results are presented in Section~\ref{section:results}, and the conclusion and future directions are drawn in Section~\ref{section:conclusion}. We also evaluate the Smagorinsky physics-based model under two coefficient settings ($C_S = C_C = 0.1$ and $C_S = C_C = 0.001$) to establish a broader benchmark for the CNN surrogate. In addition, while the integration of the developed surrogate model with the StellarBox code is out of scope of the current paper, we still assess the generalization of the trained CNN model when applied to low-resolution simulation inputs under both coefficient regimes in Section~\ref{section:results} by comparing the resulting distributions of the Reynolds stress tensor components.


\section{Background}\label{section:Background}

\subsection{StellarBox Radiative Hydrodynamic Simulations}

The StellarBox code solves the compressible MHD equations on a three-dimensional Cartesian grid, as well as a fully-coupled radiative transfer in the local thermodynamic equilibrium (LTE) approximation. As was mentioned above, the code incorporates a dynamic Smagorinsky model for SGS treatment~\citep{Smagorinsky1963_viscosity, Germano1991_dynamicSmagorinsky}. StellarBox has previously been successfully applied to model and study a variety of magnetoconvection, photospheric, chromospheric, and coronal phenomena \citep{Jacoutot2008ApJ...684L..51J,Kitiashvili2011,Kitiashvili2012ApJ...751L..21K,Kitiashvili2023MNRAS.518..504K,Sadykov2021ApJ...909...35S}. We note here that, while StellarBox is the radiative MHD code in nature, the magnetic field has been omitted in the simulation run considered in this work. Following \citep{Wray2015_stellarbox}, the key equations that are solved by the StellarBox code in the absence of a magnetic field become:

\begin{gather}
    \partial{}_{t}\rho{} + (\rho{}u_{j})_{,j} = 0 \label{eqn:sb:mass}\\
    \partial{}_{t}(\rho{}u_{i}) + (\rho{}u_{i}u_{j} + p\delta{}_{ij})_{,j} = \Pi{}_{ij,j} - \rho{}\phi{}_{,i} - 2\epsilon{}_{ijk}\Omega{}_{j}\rho{}u_{k} \label{eqn:sb:momentum}\\
    \partial{}_{t}E + [(E+p)u_{j}]_{,j} = -(\phi{}_{j}u_{i})_{,j} + (\Pi{}_{ij}u_{i})_{,j} - Q_{j,j} - Q^{rad}_{j,j} \label{eqn:sb:energy}\\
    E = \rho{}e + \dfrac{1}{2}\rho{}u_{i}u_{j} + \rho{}\phi{} \label{eqn:sb:intenergy},
\end{gather}

where $\partial{}_{t}$ represents the partial time derivative, the subscript $_{,j}$ is the partial spatial derivative in the $j$th direction, $\Pi$ is a viscous tensor, $Q$ is the non-radiative heat flux containing diffusive and Joule heating, and $Q^{rad}_{j}$ is the radiative heat flux, and $e$ is the energy per unit mass.

The StellarBox model used in this study is the same as was considered earlier in \citep{Kitiashvili2015} and \citep{Waidele2023ApJ...949...99W}. In brief, the simulated data are in the format of 512$\times$512$\times$512-size data cubes, with a spatial resolution of 12.5\, km in the horizontal scale (and a comparable uniform vertical scale), and a cadence of 30\,s. The simulated data cubes cover about 1\, Mm of the atmosphere above the geometrical $h=0\,km$ height, and $\sim$5.4\, Mm of the subphotospheric region. The simulations have been initialized with the standard solar model with small random velocity perturbations added ($\pm$1–2\,cm/s) and run until a relaxed statistically stationary state of solar convection is reached \citep{Kitiashvili2015}.

\subsection{Physics-based Subgrid Scale Models}

In subgrid-scale turbulence modeling, the Reynolds stress tensor is a key quantity of interest, representing the correlation of the small-scale velocity field that produces additional terms in the hydrodynamic / MHD equations and can lead to kinetic energy dissipation and momentum transport. Accurate modeling of the Reynolds stress tensor is essential for capturing the dynamics of turbulence. It is defined as:
\begin{gather}
    \tau{}_{ij} = \widetilde{u_{i}u_{j}} - \tilde{u_{i}}\tilde{u_{j}},
    \label{eqn:ReynoldsStress}
\end{gather}
where $u_{i}$ is the $i$-th component of the velocity, $\widetilde{u_{i}u_{j}}$ represents the average of the product of two components, and~$\tilde{u_{i}}$ indicates the averaged value of the variable, both over the spatial scale of interest. We note here that very often the $u_{x}$ component of the velocity is noted in the simulation approaches as $u$, $u_{y}$ component~--- as $v$, and $u_{z}$ component~--- as $w$. We will follow these notations in the current work.

Turbulence in the solar convection zone is anisotropic and inhomogeneous in nature \citep[e.g.,][]{Nordlund1985}. Traditional methods for estimating the Reynolds stress tensor in the convection zone rely on the spatial derivatives and statistics of the macroscopic flows, such as the Gradient model \citep{Karpov2022}, Smagorinsky model \citep{Smagorinsky1963_viscosity}, and Dynamic Smagorinsky model \citep{Germano1991_dynamicSmagorinsky}. We utilize two physics-based subgrid models as the baseline models for this study: the Gradient model and the Smagorinsky model. 

The Gradient model estimates subgrid stress based on the product of the first spatial derivatives of resolved scales. In essence, $\tau{}_{ij}$ is represented by the quadratic term of the Taylor series expansion of the resolved-scale velocities \citep{Khani2022gradient}. The model has previously been utilized as a physics-based baseline to be compared with the ML-driven turbulence models for supernovae explosion \citep{Karpov2022}. The gradient model is described as:
\begin{gather}
    \tau{}_{ij} = \dfrac{\widetilde{\Delta}^{2}}{12}\partial{}_{k}\widetilde{u_{i}}\partial{}_{k}\widetilde{u_{j}},
    \label{eqn:gradiendmodel}
\end{gather}
where $\widetilde{\Delta}$ represents the characteristic scale of the simulation grid or an averaging kernel, and $\widetilde{u_{i}}$ are the plasma velocity components averaged over these scales. Correspondingly, the $\tau{}_{ij}$ can be estimated from the directly simulated scales using the gradient model. 

The Smagorinsky model~\citep{Smagorinsky1963_viscosity} is a widely used model that assumes eddy viscosity proportional to the local strain rate. The model performs well for both homogeneous and channel-type flows, suggesting applicability to more complex flows \citep{Meneveau1994ALD}. Previously, it was demonstrated that the utilization of the Smagorinsky model in simulations of the solar convection allows for capturing energy transport on subgrid scales \citep[e.g,][]{Kitiashvili2013}. However, accurate modeling of sub-grid energy transport requires the determination of SGS coefficients, which vary for different computational setups. The typical approach to determining these coefficients is to run the dynamic Smagorinsky model~\citep{Germano1991_dynamicSmagorinsky}, which adapts the model parameters based on local flow conditions. We will be utilizing the ordinary (non-dynamic) version of the model within this study as a baseline.

Originally introduced in \citep{Smagorinsky1963_viscosity}, the aforementioned model approximates the subgrid stress tensor as:
   \begin{gather}
       \tau{}_{ij} = -\dfrac{\Pi{}_{ij}}{\widetilde{\rho}} = -2\mu{}_{T}'\left(S_{ij}-\dfrac{1}{3}\partial{}_{k}\widetilde{u_{k}}\delta{}_{ij}\right) + k_{T}'\delta{}_{ij} \\
       \mu{}_{T}' = \widetilde{\Delta}^{2}\left(C_{S}|S|+C_{D}|\partial{}_{k}\widetilde{u_{k}}|\right),~~~~~ k_{T}' = \dfrac{2}{3}C_{C}\widetilde{\Delta}^{2}|S|^{2}\\
       S_{ij} = \dfrac{1}{2}\left(\partial{}_{j}\widetilde{u_{i}} + \partial{}_{i}\widetilde{u_{j}}\right),~~~~~ |S| = \sqrt{2S_{ij}S_{ij}},
   \end{gather}

where $\Pi{}_{ij}$ represents the $ij$th component of the viscous tensor from the original StellarBox equations \ref{eqn:sb:mass}-\ref{eqn:sb:intenergy}, and $\widetilde{\rho}$ represents the average density in the cell where $\Pi$ is estimated. We also note that the StellarBox code \citep{Wray2015_stellarbox} utilizes the compressible formulations for the Smagorinsky model and has two coefficients $C_{C}$ and $C_{S}$, as well as a shock-capturing coefficient $C_{D}$.  In this study, we evaluate the Smagorinsky model under two coefficient settings: $C_S=C_C=0.1$ and $C_S=C_C=0.001$. The selection of $C_S=C_C = 0.1$ follows the suggestions in \citep{Germano1991_dynamicSmagorinsky}; in this regime, the model applies a relatively strong eddy viscosity. At $C_S = C_C = 0.001$, the SGS dissipation is substantially reduced; this set of coefficients has been utilized to run the high-resolution StellarBox simulations. The $C_{D} = 0$ has been assumed in both cases. Overall, we assess how the ML surrogate performs relative to the physics-based model across two distinct dissipation regimes. This provides a broader and more informative benchmark than comparison against a single coefficient setting.

\section{Methodology}\label{section:Methodology}


\subsection{Data Preparation}\label{section:data_prep}

In order to have labeled data for our training and test, we must first extract and label the data from existing StellarBox simulation output. As mentioned above, we considered the dataset of 3D radiative hydrodynamic simulations with the StellarBox code that were previously used in \citep{Kitiashvili2015}. Because of the high resolution of $\sim$12.5\,km, we can use these simulations to compute the Reynolds stress tensors directly (i.e., we do not assume any additional contribution from the Reynolds stresses at the subgrid-scale of this high-resolution model).

The characteristic timescale at the solar photosphere is $\sim 5-10$ minutes. Therefore, we used every 10th data cube (or a 5-minute cadence) in the data series, a total of 25 data cubes, which allows us to make each considered data cube less dependent on the others. We then slice each data cube into 4$\times$4$\times$4 sub-cubes, effectively decreasing the resolution 4 times (and making it corresponding to $\sim$50\,km). For each 3rd sub-cube in each direction, we compute the Reynolds stress tensor components directly following Eqn.~\ref{eqn:ReynoldsStress}. We also compute the average density in this sub-cube, and average velocities in this and its 26 neighboring sub-cubes. The resulting task is to predict the $\tau{}_{ij}$ components in the central sub-cube based on their density and surrounding velocity fields. Because of the non-overlapping spatial sampling and sufficiently large temporal sampling, we ensure that our dataset consists of non-correlated data points, and therefore could be split into train-validation-test datasets randomly. The final dataset has $\sim$1.8\,M individual data points sampling the upper solar convection zone and a lower photosphere.

\begin{figure}[htb!]
    \centering

    \gridline{
  \fig{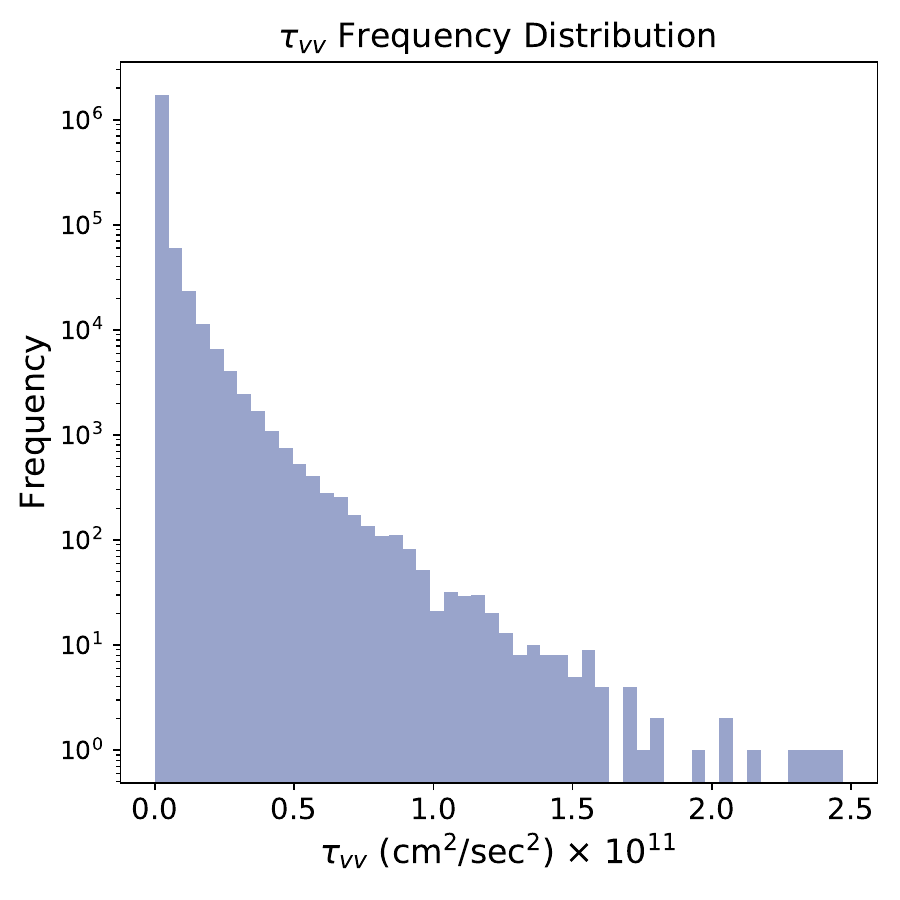}{0.48\textwidth}{\bfseries(a)}
  \fig{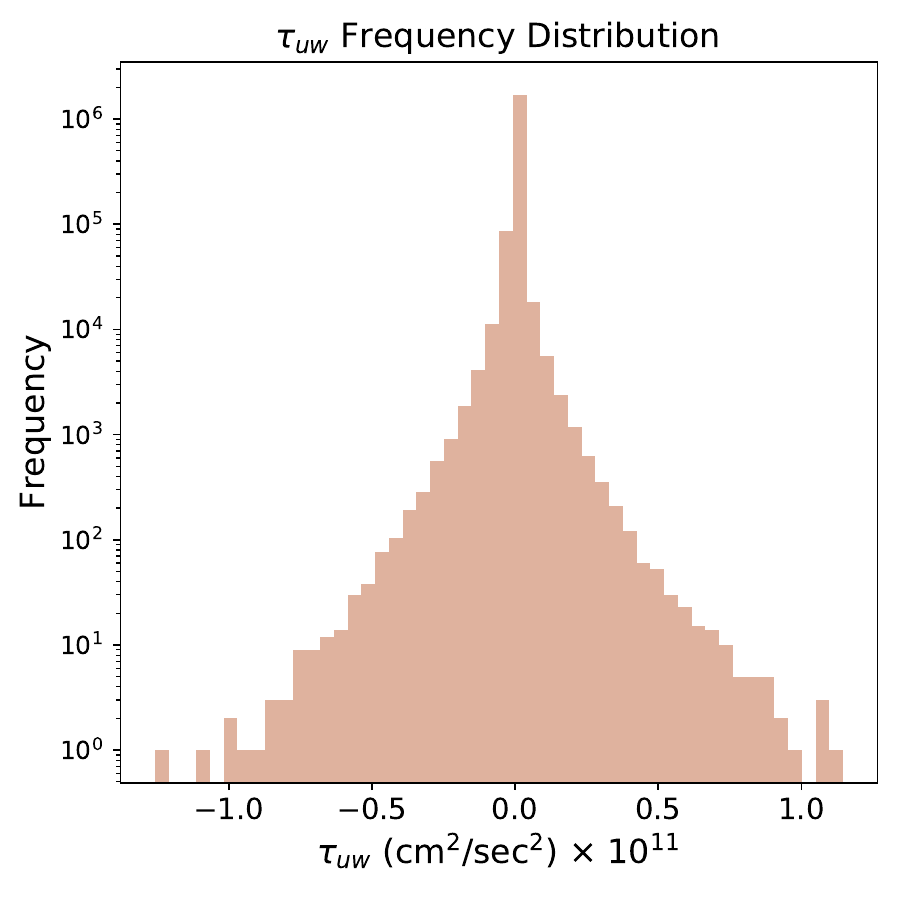}{0.48\textwidth}{\bfseries(b)}
}


    \caption{
        Panels (a) and (b) show the original distribution of $\tau_{vv}$ and $\tau_{uw}$.  All distributions are log-scaled on y-axis for better visualization.}
        \label{fig:targes_pre}
\end{figure}

\begin{figure}[htb!]
    \centering
        \gridline{
  \fig{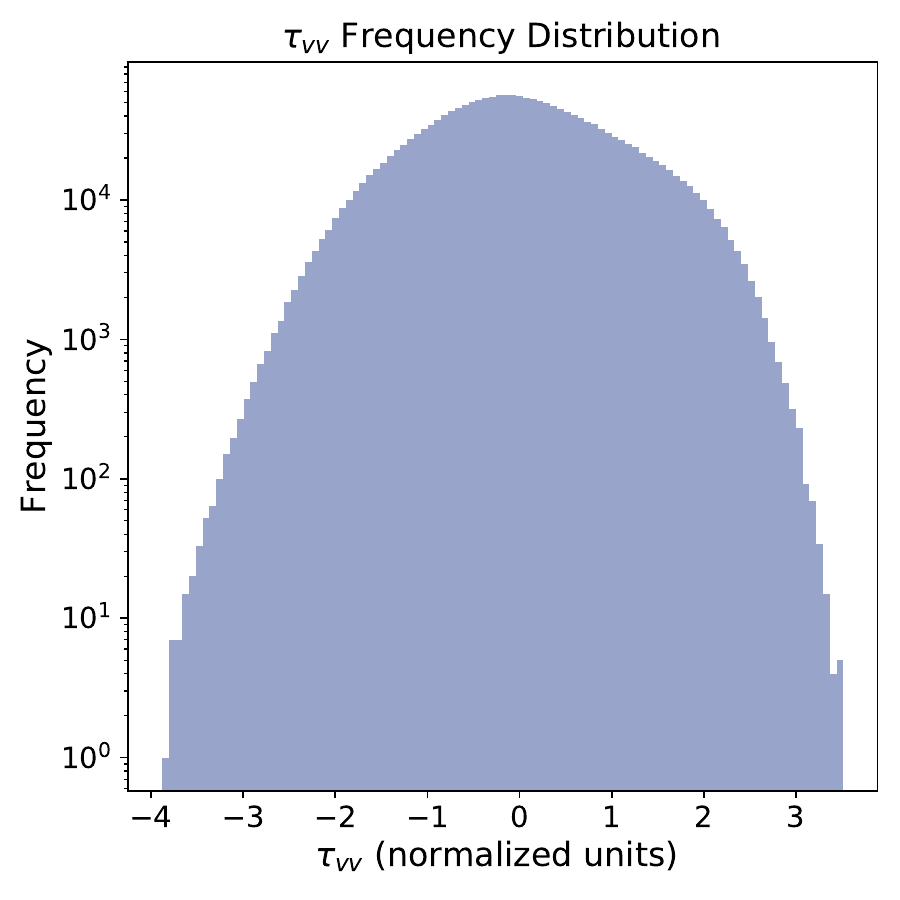}{0.48\textwidth}{\bfseries(a)}
  \fig{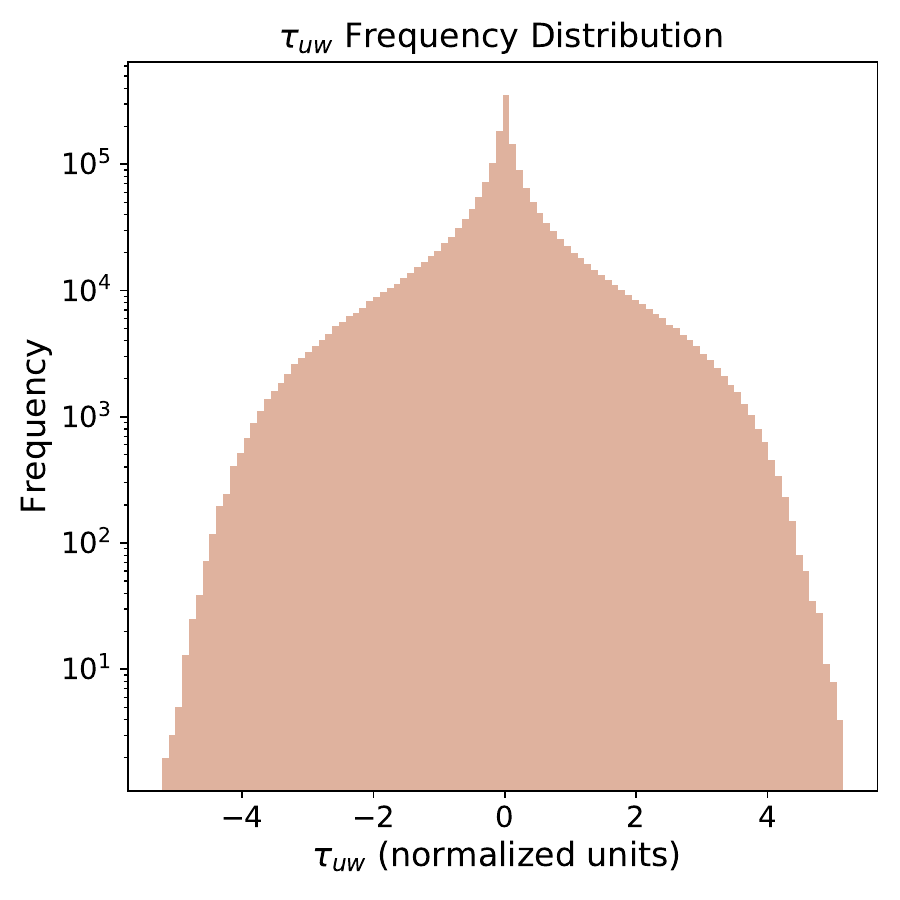}{0.48\textwidth}{\bfseries(b)}
}
    \caption{
        Panels (a) and (b) present the transformed distributions of $\tau_{vv}$ and $\tau_{uw}$ after preprocessing, respectively. Here, normalized units denote that the stresses have been log or signed-log transformed and then standardized. There was additional non-dimensionalization.
    }
    \label{fig:targets_post}
\end{figure}


The preprocessing stage is crucial for preparing the input and target variables extracted from the dataset for effective utilization in machine learning models. We begin by applying a Z-score normalization to all the velocity inputs (27 sub-cubes $\times$ 3 components) and density (one scalar) input using the `StandardScaler' \citep{scikit-learn}, which transforms each feature to have zero mean and unit variance. By performing this normalization, we ensure that each feature contributes equally to the analysis, preventing features with naturally larger scales from dominating the importance of the model’s decision-making process. 

Next, we turn our attention to the target features (Reynolds stress tensor components, $\tau{}_{ij}$), and as seen in Figure~\ref{fig:targes_pre}, the population is highly concentrated around zero, following an exponential distribution with a long tail. This leads to the assumption that a data transform would benefit the model training by making the data more normally distributed, as can be seen in Figure~\ref{fig:targets_post} showing $\tau{}_{vv}$ and $\tau{}_{uw}$ after the preprocessing steps. The following steps highlight the preprocessing pipeline for the target variables: 
\begin{enumerate}[leftmargin=*]
\item \textbf{Logarithmic Transformation}: In cases of highly skewed data with long tails, regression analysis is easier to perform using the logarithmic transformation of the data \citep{Colej3683}. 
Various studies \citep{OLIVIER2008333,bellego2022, Cleophas2016} show that logarithm transformations aim to stabilize variances and foster linear relationships between variables, thereby improving model performance. Therefore, a natural logarithm is applied to diagonal terms $\tau{}_{uu}$, $\tau{}_{vv}$, and $\tau{}_{ww}$ to address skewed distributions and promote a more uniform distribution density of the data. In Figure~\ref{fig:targes_pre}(a), we show how the distribution of the $\tau{}_{vv}$ variable looks prior to transformation. All the diagonal terms $\tau{}_{uu}$, $\tau{}_{vv}$, and $\tau{}_{ww}$ follow a similar distributions.

\item \textbf{Signed Logarithmic Transformation}: The off-diagonal target variables $\tau{}_{uv}$, $\tau{}_{vw}$, and $\tau{}_{uw}$ undergo a signed logarithmic transformation process. This method involves extracting the sign of the original values to preserve their directionality. Original values are scaled down by dividing them by a constant factor of $10^8$, bringing the data into a manageable numerical range for the logarithmic operation. A logarithmic transformation is then applied to their absolute values, with a constant offset of 1 added to avoid negative values. Finally, the original sign is applied to the transformed values. This approach effectively handles heavily zero-centered data with extreme outliers in both the positive and negative directions, as seen in the original distribution in Figure~\ref{fig:targes_pre}(b), while retaining the directionality and relative magnitudes of the variables. The scaling step was not required for strictly positive target variables, as the logarithmic transformation could be applied directly without risking undefined operations.
\item \textbf{Standard Scaling}: All preprocessed target variables are then standardized using `StandardScaler' \citep{scikit-learn}. This procedure scales each variable independently by subtracting the mean and dividing it by the standard deviation. Standard scaling ensures that the variables are on a comparable scale, preventing dominance by variables with larger magnitudes and facilitating convergence during model training. The Figure~\ref{fig:targets_post} shows the post-transformation distributions of $\tau_{vv}$ and $\tau_{uw}$, indicating more uniform distributions with respect to those for non-scaled parameters. 
\end{enumerate}

Throughout this paper, references to ``normalized units'' for Reynolds stresses (e.g., in Figures~\ref{fig:targes_pre} and \ref{fig:targets_post}) denote values that have been log or signed-log transformed and then standardized. The normalization efforts contribute to enhanced model performance, robustness, and interpretability, which we will show in Section~\ref{section:transform_effects}.

\subsection{Input Field Distributions: High vs Low Resolution}\label{section:lowres:inputdist}
Because the ML surrogate is trained using filtered high-resolution simulations, it is important to verify that the resolved-scale input variables encountered in genuine low-resolution simulations remain consistent with the statistical properties of the training data. If the low-resolution simulations produce input distributions substantially outside the training domain, the surrogate predictions may become unreliable. To evaluate whether the trained surrogate operates within a consistent input domain, we therefore compare the distributions of the resolved-scale input variables between the high-resolution training dataset and independent low-resolution ($\sim$50\,km)  simulations. Two independent low-resolution StellarBox simulations are considered, corresponding to the Smagorinsky coefficient settings introduced in Section~\ref{section:Background}: $C_S = C_C = 0.1$ and $C_S = C_C = 0.001$. For each case, we examine the distributions of the averaged velocity components ($u$, $v$, $w$) and the central cell density $\rho$.
\begin{figure*}[tb]
    \centering
    \includegraphics[width=0.99\textwidth]{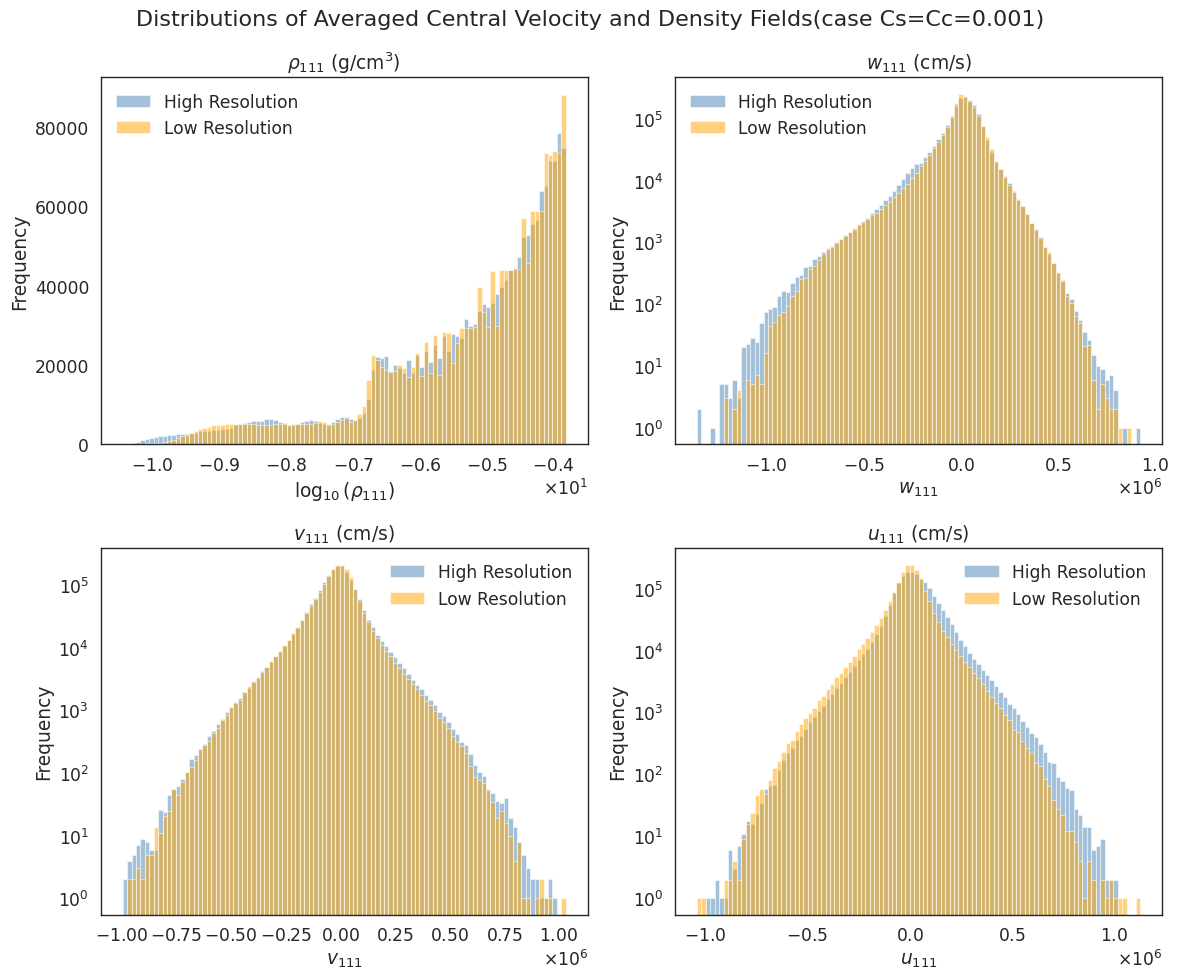}
    \caption{Distributions of Central Velocity and Density Fields when Smagorinsky coefficients are $C_S = C_C = 0.001$.}
    \label{fig:Data0001}
\end{figure*}
\begin{figure*}[tb]
    \centering
    \includegraphics[width=0.99\textwidth]{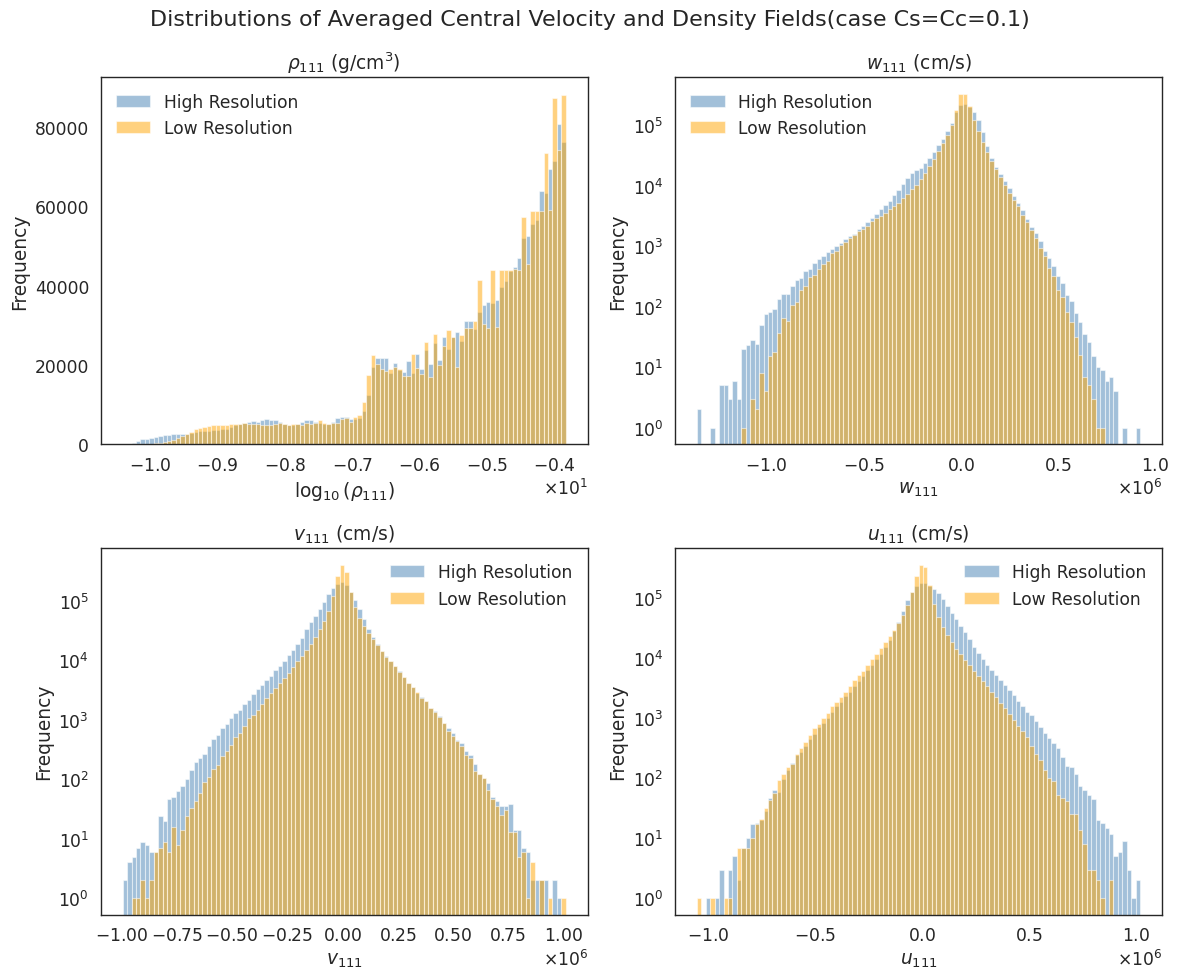}
    \caption{Distributions of Central Velocity and Density Fields when Smagorinsky coefficients are $C_S = C_C = 0.1$.}
    \label{fig:Data01}
\end{figure*}

Figure~\ref{fig:Data0001} shows the distributions of the velocity components and the density for the $C_S = C_C = 0.001$ case, and Figure~\ref{fig:Data01} shows the corresponding distributions for $C_S = C_C = 0.1$. In both cases, the high- and low-resolution distributions share the same general shape. However, despite this overall agreement, some systematic differences are apparent. The low-resolution velocity distributions are slightly narrower than their high-resolution counterparts, with reduced occurrence of high-magnitude velocities in the tails. This is expected: averaging over a coarser grid smooths out small-scale velocity fluctuations. The density distributions remain closely matched between resolutions. 
\begin{figure*}[tb]
        \centering
  \includegraphics[width=0.99\textwidth]{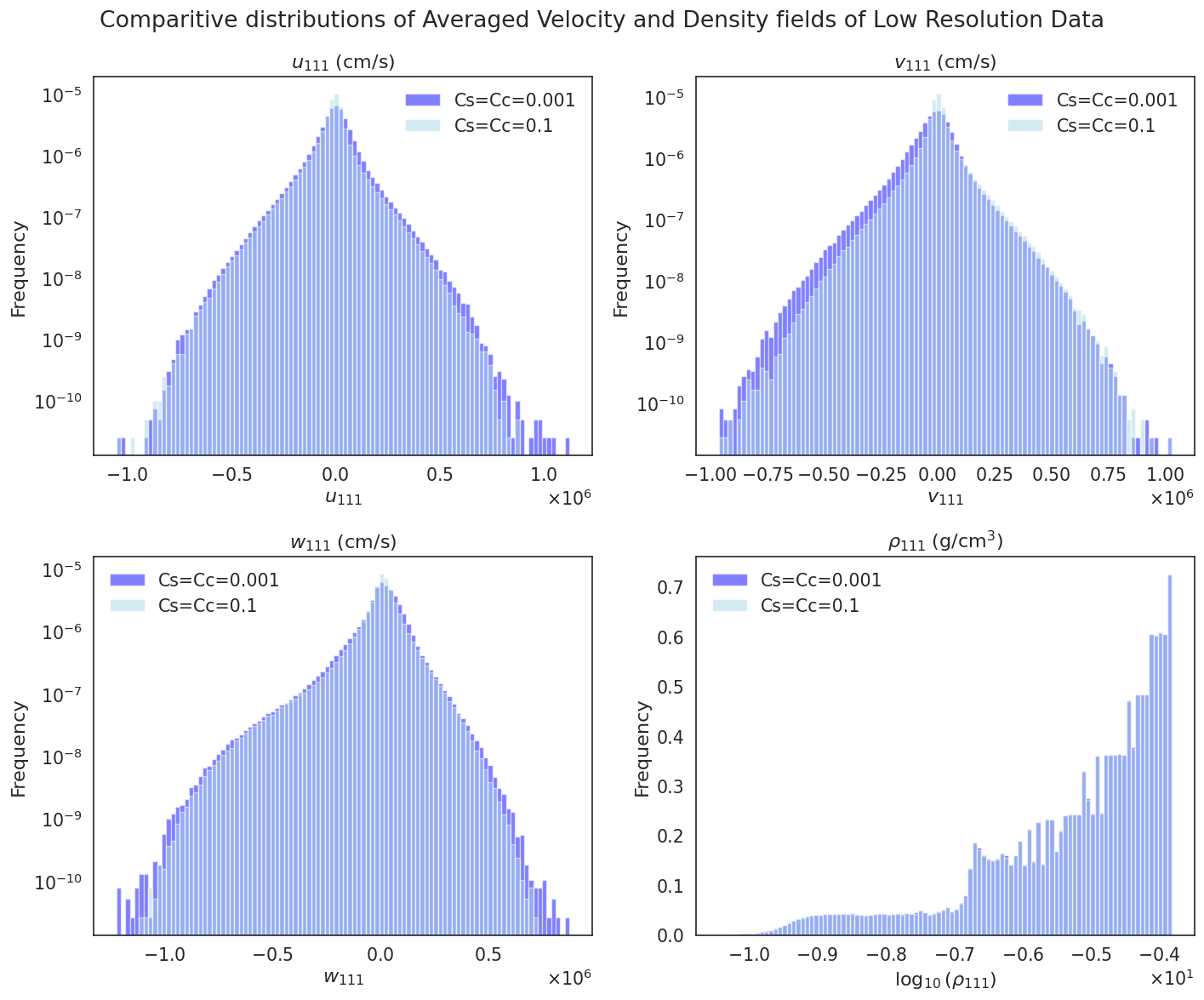}
    \caption{Comparison of velocity and density distributions for the two low-resolution simulation datasets.}
    \label{fig:lowresdata}
\end{figure*}
Comparing the two low-resolution runs against each other in Figure~\ref{fig:lowresdata}, the velocity distributions show small but discernible differences, particularly in the tail regions of $u$ and $w$. The $C_S = C_C = 0.001$ run produces slightly broader velocity distributions than the $C_S = C_C = 0.1$ run. These differences are modest but physically consistent, and they propagate into the CNN's predictions as discussed below.
\subsection{Model Description}\label{section:models}
\subsubsection{Multilayer Perceptron}
 As a simplest-case ML model, we utilize the MLPRegressor of scikit-learn \citep{scikit-learn}, a neural network designed for regression tasks. The Multilayer Perceptron (MLP) is a fundamental model in neural networks, consisting of multiple layers of fully connected neurons. Each neuron applies a non-linear activation function to a weighted sum of inputs. MLPs are particularly well-suited for regression tasks with non-linear relationships, making them a viable candidate for predicting the complex dynamics of the Reynolds Stress Tensor components. The MLPRegressor model settings are listed in Table~\ref{table:model_settings}. However, the scikit-learn implementation of MLPs can be limited in scalability and customization, leading us to explore Convolutional Neural Networks (CNNs) as a more flexible alternative, as discussed in the next section.
 
\subsubsection{Convolutional Neural Networks}\label{section:model:cnn}
\begin{figure}[htb]
    \centering
    \plotone{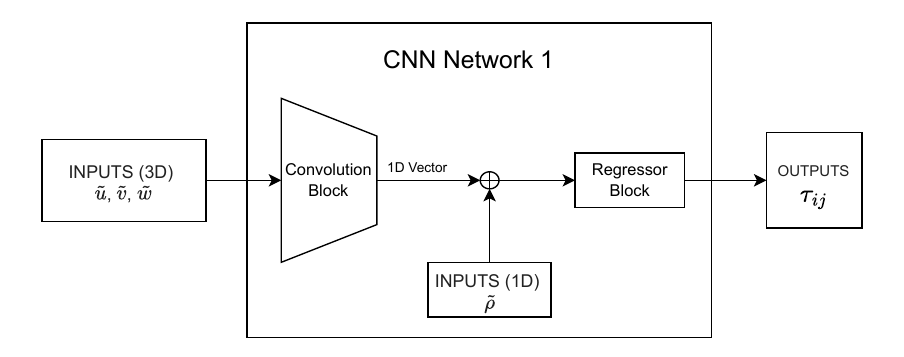}
    \caption{Model schematics of 3DCNN1.}
    \label{fig:CNN1Arch}
\end{figure}

Convolutional Neural Networks (CNNs) have become a powerful tool to capture spatial hierarchies in data through convolutional layers \citep{yamashita2018cnn_radiology}, making them suitable for tasks involving spatially correlated inputs, such as the 3D velocity fields in our study. We designed our models utilizing the conventional 3D CNN in the PyTorch library \citep{paszke2019pytorchimperativestylehighperformance}, which is an open-source machine learning library developed by Facebook's AI Research lab (FAIR). The convolutional layer, implemented using Conv3d, is a fundamental component of our model, responsible for processing spatial information from the input tensor. We investigated two different three-dimensional convolutional neural networks
(3DCNN) model designs, 3DCNN1 and 3DCNN2, differ primarily in how they incorporate the integration of the scalar density field along the velocity components. Specifically:

\textbf{3DCNN1: } This architecture processes the velocity components and a scalar density through separate pathways. The input to the model consists of a $3\times3\times3$ data cube of averaged velocity components $u,v, w$ represented as three channels along with a scalar density feature ($\rho$) located at the central sub-cube. The model design shown in Figure~\ref{fig:CNN1Arch} has two main parts: a convolutional feature extraction block and a fully connected regression block. In the convolutional block, the three input velocity channels are first processed through a series of 3D convolutional layers to extract spatial features. Following the convolutional operations, the data are flattened to transition from multi-dimensional feature maps to a single-dimensional feature vector. This separation approach allows the model to focus on spatial dependencies before integrating scalar inputs. Subsequently, the scalar density feature is concatenated with the 1D feature vector before entering the fully connected regression layers of the regression block to predict the tensor components. This concatenation integrates critical scalar information with the learned spatial features, enriching the model's input just before the final regression steps. 

In the development of 3DCNN1, multiple activation functions and layer configurations were explored to optimize the model’s ability to capture spatial hierarchies from local 3D velocity fields. In all three 3DCNN1 variants, the velocity components are first processed through two 3D convolutional layers; the first convolutional layer uses 32 filters equipped with a kernel $3\times3\times3$, with padding and stride set to 1. This preserves the spatial resolution of the input. The second convolutional layer uses 64 filters with the same kernel size, stride, and padding. 

The first model, 3DCNN1.1, uses Leaky Rectified Linear Unit (LeakyReLU) activations after each convolutional layer. LeakyReLU maintains a small non-zero gradient for negative inputs (slope $\alpha=0.01$), preventing the dying neurons problem without adding significant computational complexity \citep{maas2013rectifier, he2015delving} during training. This slope ($\alpha$) should not be confused with the L2 regularization coefficient alpha used in other contexts.
The second model, 3DCNN1.2, uses hyperbolic tangent (Tanh) activations in the convolutional layers to capture complex non-linear relationships, which outputs values in the range (-1, 1). This smooth, bounded activation might be beneficial for capturing complex, nonlinear patterns in the early stages of training. The third model, 3DCNN1.3, incorporates ReLU activations with Batch Normalization to stabilize learning and reduce internal covariate shifts, enabling deeper network training with smoother convergence.

The fully connected layers primarily use SoftSign activation. Similar to Tanh, SoftSign provides a non-linear curve, but it converges asymptotically, which can be advantageous for layers closer to the output where we need a gentle decision boundary delineation. It is effective in preventing vanishing gradient \citep{Szandala2021, app10051897} and dealing with outliers or extreme values because of its asymptotic nature, and it offers smooth gradient transitions, particularly for output layers requiring fine-tuned decision boundaries, except in 3DCNN1.3, where ReLU is used. A final dense layer with six neurons corresponding to the six components of the Reynolds stress tensor represents the output for all models.
As described in Section~\ref{section:cnn_tuning}, the key hyperparameters were tuned to balance model performance with computational efficiency and are listed in Table~\ref{table:cnn1_layers}. 

\begin{figure}[htb]
  \plotone{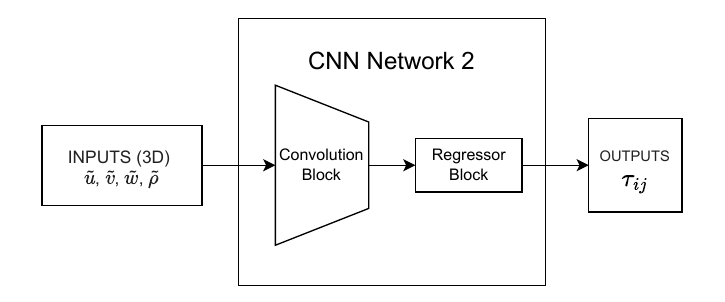} 
  \caption{Model schematics of 3DCNN2. }
  \label{fig:CNN2Arch}
\end{figure}

\textbf{3DCNN2: }
As in our 3DCNN1 model design, the input to the model consists of a $3\times3\times3$ data cube of averaged velocities with three channels corresponding to the velocity components $u$, $v$, and $w$, and one scalar feature representing the density at the central sub-cube.
However, in this model, the scalar density field is treated as an additional `channel' of identical values within the input data cube, resulting in a 4-channel input. Three channels represent the 3D velocity components ($u$, $v$, $w$), and the fourth channel corresponds to the density, $\rho$. The scalar density is reshaped and expanded spatially to match the dimensions of the velocity inputs before being concatenated. This integrated input is then jointly processed through the convolutional layers. We present the general network flow of the 3DCNN2 design in Figure~\ref{fig:CNN2Arch}.

In 3DCNN2 variants, multiple activation functions and layer configurations were explored to optimize the model’s ability to capture spatial hierarchies from the input. The two best models, one for each activation used, were selected based on their performance during the parameter tuning process described in Section~\ref{section:cnn_tuning}, and are listed in Table~\ref{table:model_settings_cnns}. The 3DCNN2.1 model uses two convolutional layers, with 32 filters in the first layer and 64 filters in the second, followed by LeakyReLU(slope $\alpha=0.01$) activations. The first convolution layer uses a $3\times3\times3$ kernel with padding, and the second convolutional layer reduces the kernel size to $2\times2\times2$ (with no padding), resulting in a reduction of spatial dimensions. 3DCNN2.2 follows a similar structure but replaces LeakyReLU with Tanh activations. The output from the convolutional layers is flattened into a 1D tensor, followed by a fully connected layer of 128 units with a SoftSign activation to stabilize learning. This model setup aims at potentially more complex feature extraction, at the cost of increased computational complexity. 

\section{Experimental Settings}\label{section:settings}
\subsection{Parameter Tuning MLP}\label{section:mlp_tuning}
To find the best-performing MLP implementation, we performed hyperparameter tuning utilizing the built-in GridSearchCV method of scikit-learn \citep{scikit-learn}, which is a technique used to systematically search through a predefined grid of hyperparameters and select the combination that yields the best performance. The parameters that were searched and the ranges we explored for MLP are given below. For hidden layer sizes, we tested various smaller, intermediate, and larger layer configurations like (50, 25), (60,50), (64, 32, 16), (82, 50, 25, 6), (100, 100). We evaluated two activation functions, ReLU and Tanh, to introduce non-linearity into the model. Optimization solvers such as Stochastic Gradient Descent (SGD) and Adam were compared for weight updates during training. Regularization was applied with alpha values of 0.0001, 0.001, and 0.01 to prevent overfitting. Learning rates were varied between constant and adaptive modes, with initial values set to 0.001, 0.01, and 0.1. We set a maximum number of 1000 iterations for training and used early stopping to halt the process when no further improvement was seen in validation scores.

The optimal MLP model settings found through the hyperparameter tuning GridSearchCV method are summarized in Table~\ref{table:model_settings}. By leveraging GridSearchCV, we aimed to optimize the model's architecture and training process, leading to improved performance in predicting the subgrid Reynolds stress tensor components. 

\newcommand{\rowsp}{\rule{0pt}{4.0ex}}
\newcommand{\rowspb}{\rule{0pt}{0.5ex}}
\begin{deluxetable}{ccccccc}
\tablecaption{Best model settings for MLP.\label{table:model_settings}}
\tablehead{
\colhead{\textbf{Layers}} & \colhead{\textbf{Activation}} & \colhead{\textbf{Solver}} & \colhead{\textbf{Alpha}} & \colhead{\textbf{Learning Rate}} & \colhead{\textbf{Learning Init}} & \colhead{\textbf{Max Iterations}} 
}
\startdata
(100, 100) & Tanh & Adam & $0.01$ & Adaptive & $0.001$ & $1,000$ \rowsp\\ \rowspb
\enddata
\end{deluxetable}

\subsection{Parameter Tuning on CNN}\label{section:cnn_tuning}
In the development of both CNN architectures (3DCNN1 and 3DCNN2), careful consideration was given to tuning the hyperparameters to ensure optimal performance based on training stability and model accuracy. The tuning process involved systematically varying key parameters like learning rate, batch size, number of epochs, activation functions, and optimizer types. All the models were trained using the Mean Squared Error (MSE) as the primary loss function and R-squared (R$^2$) score as a secondary performance metric. The R$^2$ metric, also known as \emph{coefficient of determination}, quantifies how much of the variance in the target variable is explained by the predictions of the model. 

Initially, the learning rate, a critical parameter for the training process, was varied. We tested learning rates from 0.0001 to 0.01 to identify the rate that allowed for quick convergence without causing the training to become unstable. After several experiments, it was observed that a learning rate of 0.001 consistently led to faster and more stable convergence. We experimented with different optimizers and selected Adam for all models due to its adaptive learning rate capabilities, which provided better convergence behavior compared to Stochastic Gradient Descent (SGD). The number of epochs, which determines how many times the training algorithm works through the entire dataset, was another significant parameter. To avoid overfitting while ensuring sufficient training, we employed early stopping with a patience of 10 epochs. The training process was set to stop if the validation loss did not improve for 10 consecutive epochs. Training typically converged by 50 epochs, beyond which no significant improvement in validation accuracy was observed. The batch size, which impacts gradient estimation in each iteration, was also optimized. We tested various sizes ranging from 32 to 256, with smaller batches tending to provide more stable but slower convergence. A batch size of 128 offered a practical trade-off, providing a robust gradient estimation while maintaining a reasonable training duration.

\begin{deluxetable*}{llllccc}
\tablecaption{Architecture settings of 3DCNN1 variants \label{table:cnn1_layers}}
\tabletypesize{\footnotesize} 
\tablehead{
\colhead{\textbf{Layer}} &
\colhead{\textbf{3DCNN1.1}} &
\colhead{\textbf{3DCNN1.2}} &
\colhead{\textbf{3DCNN1.3}} &
\colhead{\textbf{Kernel}}&
\colhead{\textbf{Stride}}&
\colhead{\textbf{Padding}}
}
\startdata
Conv Layer 1 &
\begin{tabular}[t]{@{}l@{}}Conv3D(3, 32),\\ LeakyReLU($\alpha=0.01$)\end{tabular} &
\begin{tabular}[t]{@{}l@{}}Conv3D(3, 32),\\Tanh\end{tabular} &
\begin{tabular}[t]{@{}l@{}}Conv3D(3, 32), BatchNorm3D,\\ReLU\end{tabular} &
$3\times3\times3$ & $1 $& $1$ \rowsp\\
Conv Layer 2 &
\begin{tabular}[t]{@{}l@{}}Conv3D(32, 64),\\ LeakyReLU($\alpha=0.01$)\end{tabular} &
\begin{tabular}[t]{@{}l@{}}Conv3D(32, 64),\\Tanh\end{tabular} &
\begin{tabular}[t]{@{}l@{}}Conv3D(32, 64), BatchNorm3D,\\ReLU\end{tabular} &
$3\times3\times3$ & $1 $& $1$ \rowsp\\
Flatten & Yes & Yes & Yes & \textemdash{}& \textemdash{}& \textemdash{}  \rowsp\\
FC Layer 1 &
\begin{tabular}[t]{@{}l@{}}Linear(1728, 256),\\Softsign\end{tabular} &
\begin{tabular}[t]{@{}l@{}}Linear(64, 256),\\Softsign\end{tabular} &
\begin{tabular}[t]{@{}l@{}}Linear(1728, 256),\\ReLU\end{tabular} &
\textemdash{}& \textemdash{}& \textemdash{} \rowsp \\
FC Layer 2 &
\begin{tabular}[t]{@{}l@{}}Linear(257, 64),\\Softsign\end{tabular} &
\begin{tabular}[t]{@{}l@{}}Linear(257, 128),\\Softsign\end{tabular} &
\begin{tabular}[t]{@{}l@{}}Linear(257, 64),\\ReLU\end{tabular} &
\textemdash{}& \textemdash{}& \textemdash{}  \rowsp\\
Output Layer & Linear(64, 6) & Linear(128, 6) & Linear(64, 6) & \textemdash{} & \textemdash{}& \textemdash{}\rowsp\\ \rowspb
\enddata
\end{deluxetable*}

As described in Section~\ref{section:model:cnn}, we introduced different activation functions in our CNN models, such as Tanh, LeakyReLU, and ReLU, between the convolution layers. The Tanh activation function was initially preferred for its smooth gradient and output normalization properties, which could improve early convergence. However, replacing the Tanh activation function with LeakyReLU (3DCNN1.1, 3DCNN2.1) in the convolutional layers ultimately provided a consistent gradient for positive inputs and a small gradient for negative inputs, which resulted in faster training and better convergence. The slight negative slope ($\alpha=0.01$) helped in reducing the vanishing gradients \citep{maas2013rectifier}, common in deep networks using saturating activations like Tanh (3DCNN1.2, 3DCNN2.2) and ReLU (3DCNN1.3). 
\begin{deluxetable*}{lllccc}
\tablecaption{Architecture settings of 3DCNN2 variants \label{table:model_settings_cnns}}
\tabletypesize{\footnotesize} 
\tablehead{
\colhead{\textbf{Layer}} &
\colhead{\textbf{3DCNN2.1}} &
\colhead{\textbf{3DCNN2.2}} &
\colhead{\textbf{Kernel}}&
\colhead{\textbf{Stride}}&
\colhead{\textbf{Padding}}}

\startdata
Conv Layer 1 &
\begin{tabular}[t]{@{}l@{}}Conv3D(4, 32),\\ LeakyReLU($\alpha=0.01$)\end{tabular} &
\begin{tabular}[t]{@{}l@{}}Conv3D(4, 32),\\Tanh\end{tabular} &
$3\times3\times3$ & 1 &1  \rowsp\\
Conv Layer 2 &
\begin{tabular}[t]{@{}l@{}}Conv3D(32, 64),\\ LeakyReLU($\alpha=0.01$)\end{tabular} &
\begin{tabular}[t]{@{}l@{}}Conv3D(32, 64),\\Tanh\end{tabular} &
$2\times2\times2$ & 1 & 0 \rowsp\\
Flatten & Yes & Yes & \textemdash{} & \textemdash{}& \textemdash{}\rowsp\\
FC Layer 1 &
\begin{tabular}[t]{@{}l@{}}Linear(64, 128),\\Softsign\end{tabular} &
\begin{tabular}[t]{@{}l@{}}Linear(64, 128),\\Softsign\end{tabular}  &
\textemdash{}& \textemdash{}& \textemdash{} \rowsp\\
Output Layer & Linear(128, 6) & Linear(128, 6)  & \textemdash{} & \textemdash{}& \textemdash{}\rowsp\\ \rowspb
\enddata
\end{deluxetable*}
Finally, the architecture-specific parameters, such as the number of filters in each convolutional layer and the size of the kernels, were adjusted as introduced earlier in Section~\ref{section:model:cnn}. Increasing the number of filters from 32 to 64 while keeping the kernel size at 3 provided improved performance as measured by training loss and model accuracy. The use of more filters in deeper layers did not significantly increase the computational load while allowing the model to capture spatial patterns more effectively, as also seen in \citep{he2016deep} and \citep{simonyan2014very}, especially for spatial inputs \citep{zhou2014object}. The models were trained using a Mean Squared Error (MSE) loss function, which is suitable for this regression task as it penalizes the squared differences between the predicted and target values. 

\begin{figure}[ht]
    \centering
    \includegraphics[width=\linewidth]{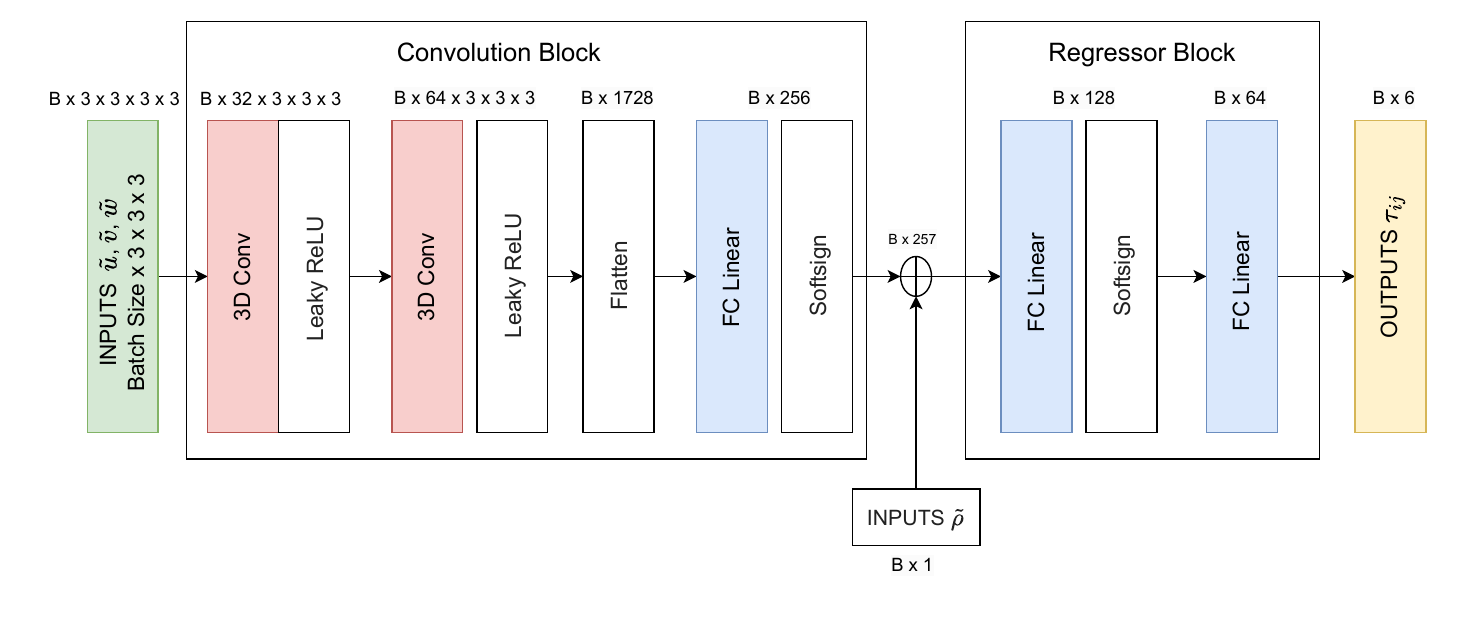}
    
    \caption{ Final CNN architecture based on the model schematics of 3DCNN1 in Figure~\ref{fig:CNN1Arch} selected for experimentation and comparison against physics-based baselines. }
    \label{fig:FinalCNNArch}
\end{figure}
The best models, one for each activation used, were selected based on their performance during the parameter tuning process and are listed in Table~\ref{table:cnn1_layers} for 3DCNN1 and Table~\ref{table:model_settings_cnns} for 3DCNN2. These models represent the best-performing configurations for their respective activation functions and layer settings, selected after extensive experimentation with filter sizes, kernel configurations, and other parameters.

\section{Discussion of Results}\label{section:results}
All quantitative results are reported on a held-out test set that was not used for training or model selection. Unless explicitly stated, statistics are computed for all samples and for all six Reynolds stress components ${\tau}=\{\tau_{uu},\tau_{vv},\tau_{ww},\tau_{uv},\tau_{uw},\tau_{vw}\}$. Training was performed using normalized targets as described in Section~\ref{section:data_prep}, but we invert all transformations prior to quantitative evaluation. The metrics used to compare models include Mean Squared Error (MSE), Root Mean Squared Error (RMSE), and the score $R^{2}$ to report their per–component values and macro–averages $\overline{\mathrm{MSE}},\,\overline{\mathrm{RMSE}},\,\overline{R}^{2}$ across the six components. The errors are reported in physical units of $\mathrm{cm}^{2}/\mathrm{s}^{2}$, and the squared errors are reported in $\mathrm{cm}^{4}/\mathrm{s}^{4}$ units. For compactness, we also report RMSE (the same units as ${\tau}$).

In this section, we extensively compare: (i) the gradient model, (ii) the Smagorinsky model with two coefficient settings ($C_S = C_C = 0.1$ and $C_S = C_C = 0.001$), (iii) the MLPRegressor model, and (iv) 3D CNN designs (3DCNN1 and 3DCNN2).

\subsection{Quantitative Performance of MLP Regressor, CNN Models, and Baselines}
The model optimized using GridSearch with the parameters listed in Table~\ref{table:model_settings} is selected as the final model for MLPRegressor. This model showed an average MSE of $6.56 \times10^{18}\,\mathrm{cm}^{4}/\mathrm{s}^{4}$ and an average explained variance (R$^{2}$) of 0.3683. 


Next, we compare the five variants of the model introduced in Section~\ref{section:cnn_tuning}. In Table~\ref{table:CNNs} we report the average R$^2$ score and the average MSE for all stress tensor target features, for each of the CNN models evaluated, with physical units restored after training.  An R$^2$ value of 1 represents a perfect fit, while an R$^2$ of 0 indicates no better performance than simply predicting the mean of the data. Among the 3DCNN1 designs, 1.2 and 1.3 exhibit a higher Mean Squared Error and a lower R$^2$ score compared to architecture 1.1 which achieves the best performance, with an average MSE of $4.89\times 10^{18}\,\mathrm{cm}^{4}/\mathrm{s}^{4}$ and an $R^2=0.54$. The weaker results of 1.2 and 1.3 indicate that larger convolutional kernels or strides decrease sensitivity to fine-scale velocity correlations. Among the 3DCNN2 designs, 2.1 performs competitively, slightly worse than 3DCNN1.1, while architecture 2.2 shows a higher MSE and a lower R$^2$ score compared to 2.1 due to oversmoothing from a larger receptive field. In general, 3DCNN1.1 performs better than all other models with the lowest average MSE of $4.89\times 10^{18}\,\mathrm{cm}^{4}/\mathrm{s}^{4}$ and $R^2=0.54$, followed by 3DCNN2.1. For subsequent comparisons, 3DCNN1.1 and 3DCNN2.1 are selected as representative CNN models of the two variants.
\begin{deluxetable}{ccc}
\tablecaption{Performance of 3D CNN variants. \label{table:CNNs}}
\tablehead{
\colhead{\textbf{Model}} & \colhead{\textbf{Average MSE}} & \colhead{\textbf{Average $R^2$}}
}
\startdata
3DCNN1.1 & $4.89\times10^{18}$ & 0.54 \rowsp\\
3DCNN1.2 & $6.07\times10^{18}$ & 0.45 \rowsp\\
3DCNN1.3 & $6.30\times10^{18}$ & 0.27 \rowsp\\
3DCNN2.1 & $5.03\times10^{18}$ & 0.53 \rowsp\\
3DCNN2.2 & $8.00\times10^{18}$ & 0.28 \rowsp\\ \rowspb
\enddata
\end{deluxetable}


   
To assess the performance of the best versions of 3DCNN1 and 3DCNN2, we compare their RMSE with each other, against the baselines, Gradient \& Smagorinsky models, and the MLP Regressor on the six targets. In Table~\ref{table:results}, we show the RMSE values of all the models for each stress tensor component, target $\tau_{ij}$. Among the physics-based baselines, the Smagorinsky model with $C_S = C_C = 0.001$ shows the highest RMSE, particularly on the diagonal terms $\tau_{uu}$, $\tau_{vv}$, and $\tau_{ww}$, where it substantially exceeds another Smagorinsky model with the $C_S = C_C = 0.1$ coefficients. We can observe that MLP shows a lower RMSE in $\tau_{uu}, \tau_{vv}$ and $\tau_{ww}$ when compared to the physics-based baselines. However, in the off-diagonal components, the RMSE does not seem to improve when compared to the Gradient model; therefore, we proceed to discuss the results of the CNN models. Compared to baseline models and MLP, 3DCNN1 and 3DCNN2 show a lower RMSE and also perform similarly on the targets $\tau_{uv}$, $\tau_{uw}$, and $\tau_{vw}$. Overall, model 3DCNN1 demonstrated the lowest RMSE on all targets, while the Smagorinsky models showed the highest RMSE on all target components. The $C_S = C_C = 0.1$ Smagorinsky model performs better than 
the $C_S = C_C = 0.001$ case on diagonal components, though both settings are outperformed by all machine learning models. This contrast between the two coefficient settings motivates the more detailed error distribution analysis presented in Section~\ref{section:errorhist}.

All versions of the 3DCNN1 and 3DCNN2 models have been trained in a maximum of 300 epochs with early stopping. It was observed that 3DCNN2s required a larger number of epochs before the early stopping criterion was met and, therefore, converged later. This is due to the increase in the dimensionality of the dataset while training 3DCNN2, as the density is expanded to a 3D vector from a single scalar value. The average MSE, RMSE, and R$^2$ scores indicate that model 3DCNN1 performs slightly better than 3DCNN2. Therefore, we consider 3DCNN1, the best CNN, for further comparisons in the next sections. The final model architecture design of (the best among 3DCNN1 family) is shown in Figure~\ref{fig:FinalCNNArch}.

\begin{deluxetable*}{ccccccc}
\tablecaption{RMSE results for the considered physics-based baseline and machine learning models. 
The RMSE measures are presented in units of cm$^{2}/$s$^{2}$. \label{table:results}}
\tablehead{
\colhead{\textbf{Method}} &
\colhead{\textbf{$\tau_{uu}$}} &
\colhead{\textbf{$\tau_{uv}$}} &
\colhead{\textbf{$\tau_{uw}$}} &
\colhead{\textbf{$\tau_{vv}$}} &
\colhead{\textbf{$\tau_{vw}$}} &
\colhead{\textbf{$\tau_{ww}$}}
}
\startdata
Gradient     & $3.53\times10^{9}$ & $1.55\times10^{9}$ & $1.81\times10^{9}$ & $3.57\times10^{9}$ & $1.83\times10^{9}$ & $4.45\times10^{9}$ \rowsp\\
Smagorinsky(Cs=Cc=0.1)  & $3.71\times10^{9}$ & $2.21\times10^{9}$ & $2.56\times10^{9}$ & $3.67\times10^{9}$ & $2.57\times10^{9}$ & $4.66\times10^{9}$ \rowsp\\
Smagorinsky(Cs=Cc=0.001)  & $4.37\times10^{9}$ & $1.85\times10^{9}$ & $2.21\times10^{9}$ & $4.39\times10^{9}$ & $2.21\times10^{9}$ & $5.69\times10^{9}$ \rowsp\\
MLP          & $2.82\times10^{9}$ & $1.66\times10^{9}$ & $1.86\times10^{9}$ & $2.88\times10^{9}$ & $2.18\times10^{9}$ & $3.50\times10^{9}$ \rowsp\\
3DCNN1 & $2.36\times10^{9}$ & $1.42\times10^{9}$ & $1.67\times10^{9}$ & $2.50\times10^{9}$ & $1.70\times10^{9}$ & $3.13\times10^{9}$ \rowsp\\
3DCNN2       & $2.46\times10^{9}$ & $1.43\times10^{9}$ & $1.71\times10^{9}$ & $2.50\times10^{9}$ & $1.71\times10^{9}$ & $3.17\times10^{9}$ \rowsp\\ \rowspb
\enddata
\end{deluxetable*}

\subsection{Comparison of PDF of Machine Learning Models: MLP and CNN}

\begin{figure*}[tb]
        \centering
        \gridline{
  \fig{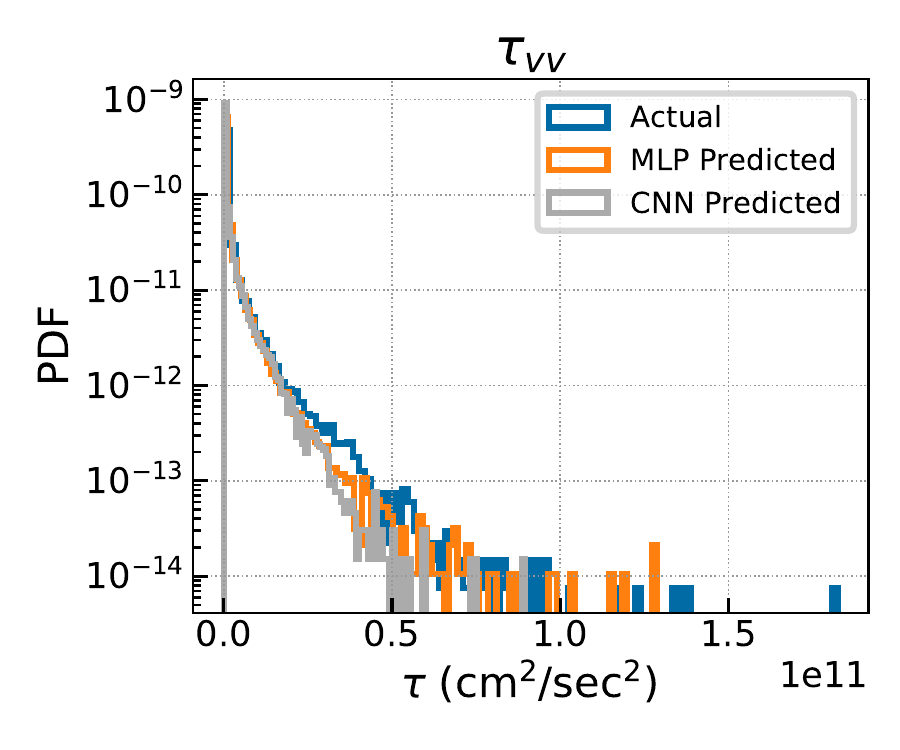}{0.48\textwidth}{\bfseries(a)}
  \fig{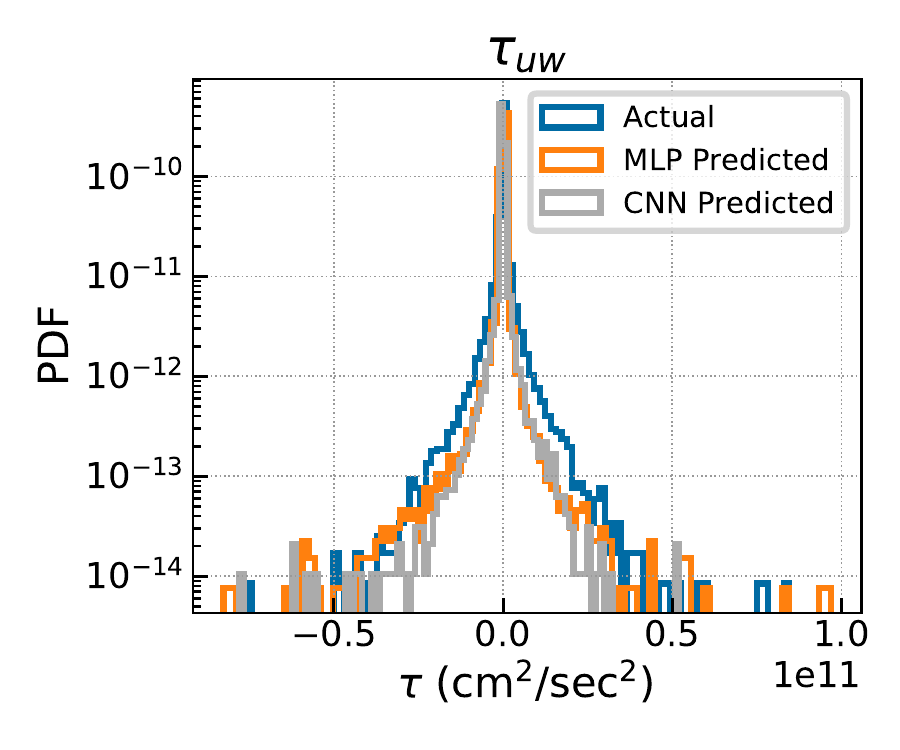}{0.48\textwidth}{\bfseries(b)}
}
    
    \caption{Distributions of components $\tau_{vv}$ shown in Panel(a) and $\tau_{uw}$ in (b). Blue curves show the original target data, orange shows the predictions of the MLPRegressor, and gray shows the predictions of the CNN.}
    \label{fig:AllPDF}
\end{figure*}

In Figure~\ref{fig:AllPDF}, we show the probability density function of the distribution of the original and predicted values of components $\tau_{uw}$ and $\tau_{vv}$ using the MLP Regressor and CNN models. For $\tau_{vv}$ in Figure~\ref{fig:AllPDF}a, the CNN predictions align well with the actual distribution near the dominant peak close to zero, resulting in lower RMSE. However, particularly in the mid-range between $0.25\times 10^{11} $cm$^{2}/$sec$^{2}$ and $0.5\times 10^{11}$ cm$^{2}$/sec$^{2}$, CNN tends to underestimate the number of points of the values, while MLP tends to preserve this part of the distribution more closely, though with greater variance in the tail regions.
Similarly, for $\tau_{uw}$ in Figure~\ref{fig:AllPDF}b, the CNN captures well the central peak and symmetrical nature of the distribution around zero but collapses the distribution in its wings, underestimating the occurrence of larger magnitude stresses. In contrast, the MLP spreads the predictions further into the distribution tails, capturing the larger or more extreme values better, although having a noisier alignment near the center and outliers near the tails. As is common in machine learning, both models struggle in areas with low numbers of samples, which constitutes a classic data imbalance issue. In general, CNN emphasizes accuracy in bulk regions, the most frequent ranges, by avoiding larger deviations in the tails, which lead to lower RMSE, while the MLP reflects a wider range of variability at the expense of noisier outliers. Therefore, we select the CNN model as the best ML model (lowest average MSE, lowest RMSE, and the highest $R^{2}$ score) for subsequent comparative analyses with the baselines, while we should always keep in mind the related drawbacks of this selection.

\subsection{Comparison of Error distributions: CNN and Baselines}\label{section:errorhist}

\begin{figure}[htb]
            \centering
        \gridline{
  \fig{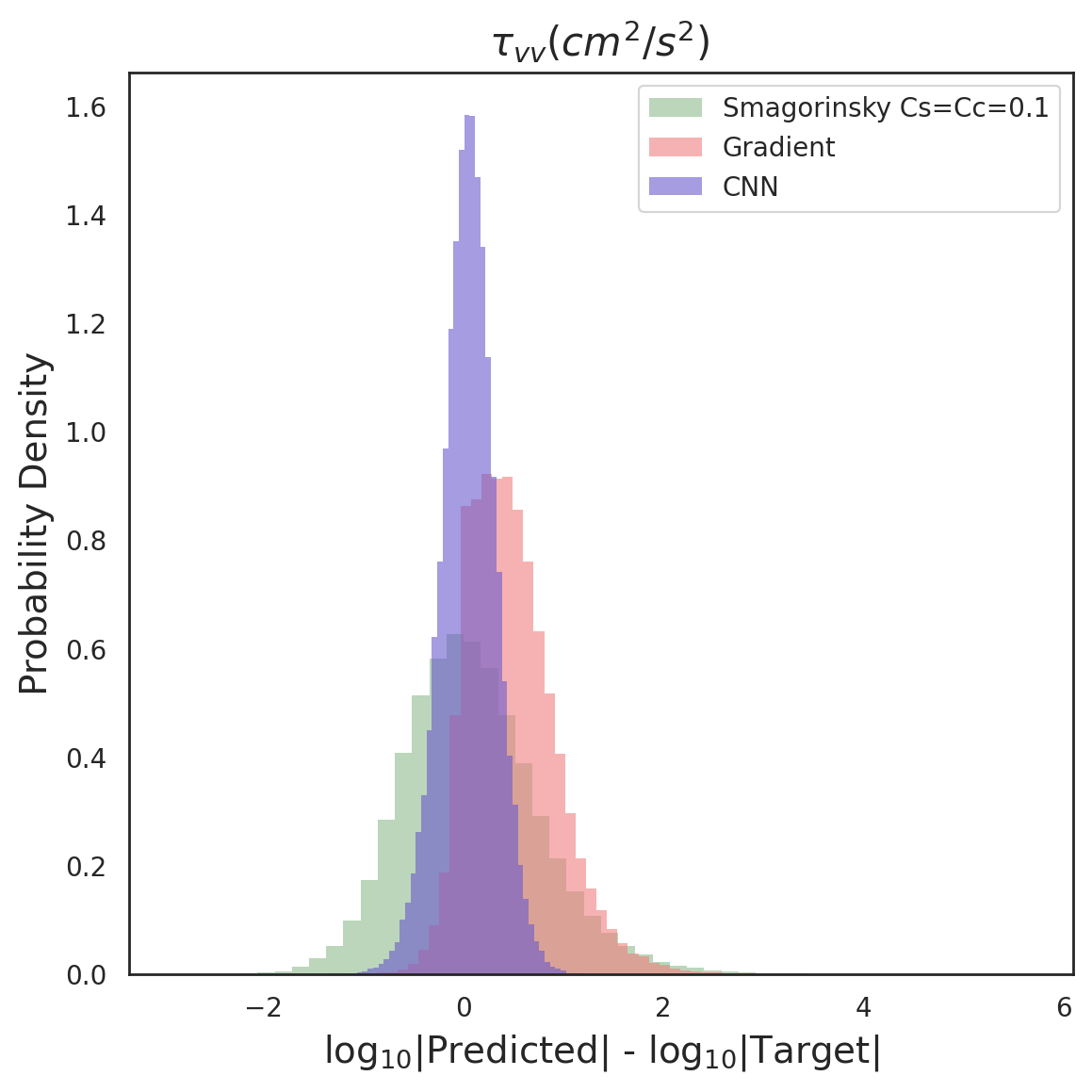}{0.48\textwidth}{\bfseries(a)}
  \fig{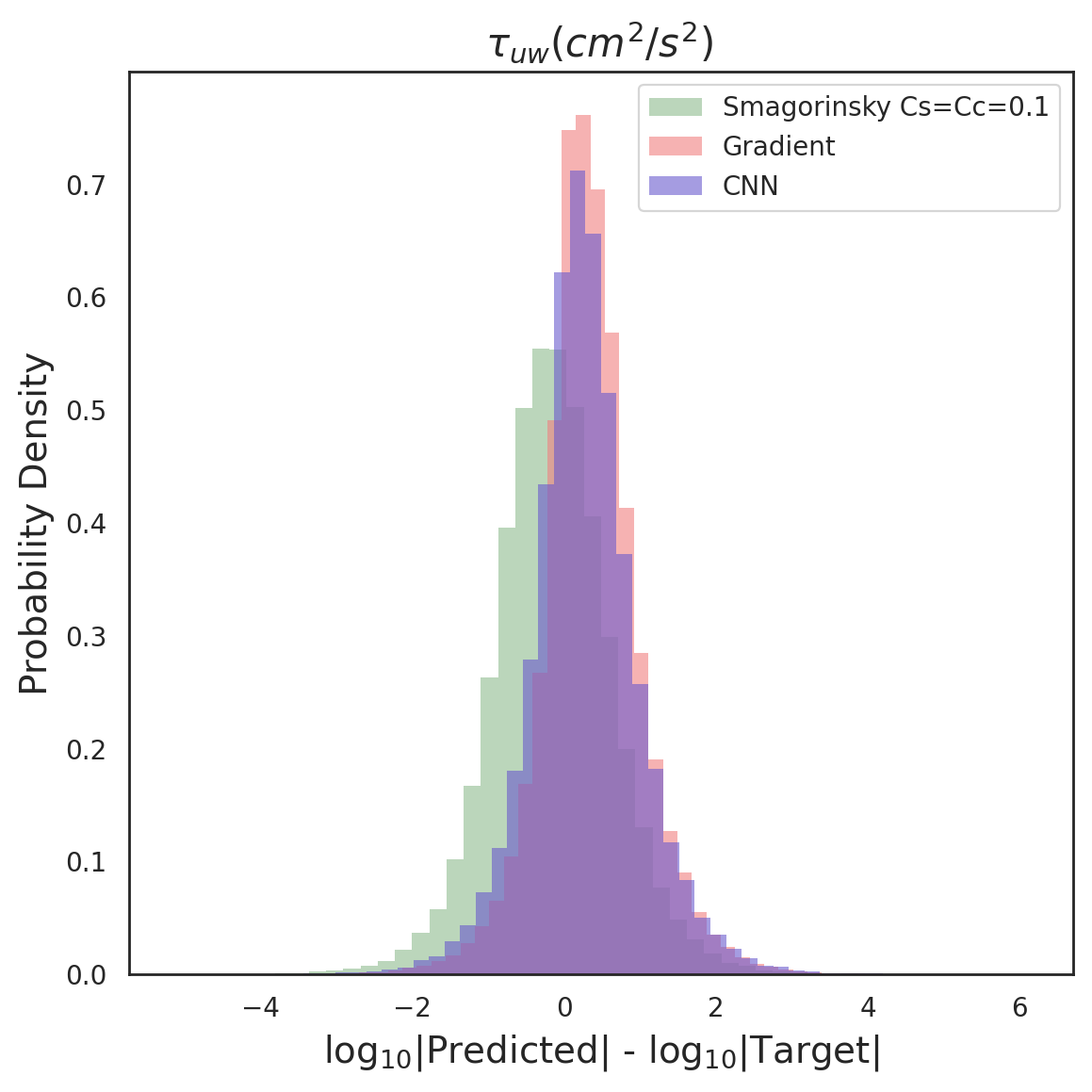}{0.48\textwidth}{\bfseries(b)}
}            

        \caption{ The Frequency plots of the deviations of $\tau_{vv}$ in Panel(a) and $\tau_{uw}$ in (b) between the predictions of the best CNN model (purple), Smagorinsky $C_S = C_C = 0.1$ (green), and Gradient models (red) and the target value.}
        \label{fig:ErrHist12}
\end{figure}

For a better understanding of the models, Figure~\ref{fig:ErrHist12} presents the distributions of the deviations between the predicted and target Reynolds stresses for the Gradient, Smagorinsky ($C_S = C_C = 0.1$), and CNN models for $\tau_{vv}$and $\tau_{uw}$. The errors are expressed as $\log_{10}(|\text{Prediction}|)-\log_{10}(|\text{Target}|)$. Correspondingly, the errors will represent the order-of-magnitude differences between predicted and true values. This log-scale formulation allows for a symmetric and scale-invariant comparison of under- and over-predictions. The histograms are normalized to show probability densities, enabling a direct comparison of the spread and the central tendency of each model's error distributions which reveal significant differences between the models. Across both  $\tau_{vv}$ and $\tau_{uw}$, the CNN model demonstrates the most concentrated error distribution around zero, indicating that the majority of the model's predictions are closely aligned with the true values. The peak of the CNN's distribution is sharply centered at zero, with fewer large errors compared to the other models. 

The Smagorinsky model (with $C_S = C_C = 0.1$) is approximately unbiased, i.e., its distribution is centered near zero, but displays the broadest distributions and heavier tails, indicating overall less accurate predictions. While it captures general trends, it lacks in precision with respect to the neural network-based models. The Gradient model shows intermediate behavior. It performs poorly for $\tau_{vv}$, with a large number of predictions deviating from the true values, as evidenced by its broad distribution and long tails in Figure~\ref{fig:ErrHist12}a. However, for $\tau_{uw}$ shown in Figure~\ref{fig:ErrHist12}b, its central peak and spread are comparable to CNN with slightly heavier tails and a mild positive shift demonstrating the tendency to overpredict.

\begin{figure}[htb]
            \centering
        \gridline{
  \fig{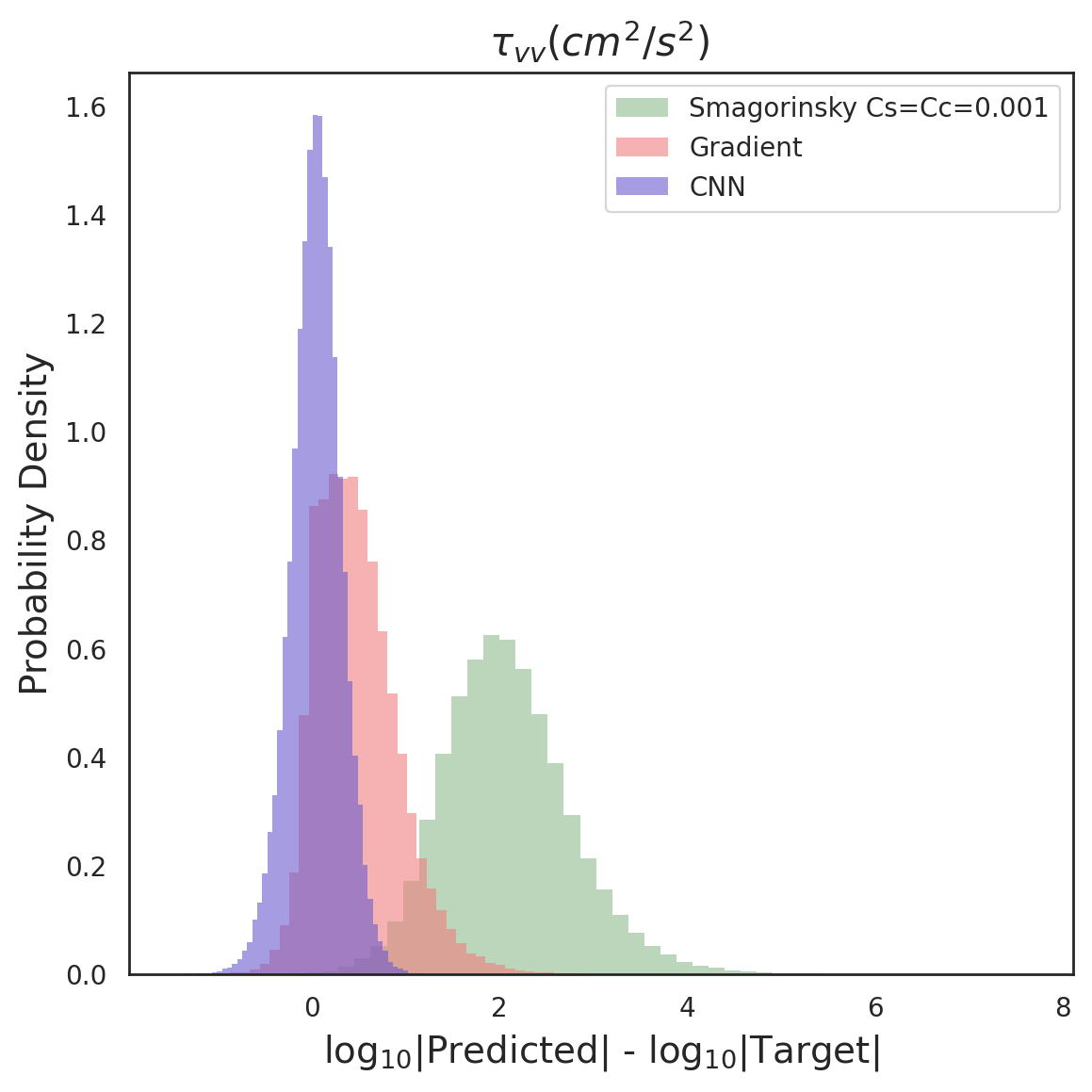}{0.48\textwidth}{\bfseries(a)}
  \fig{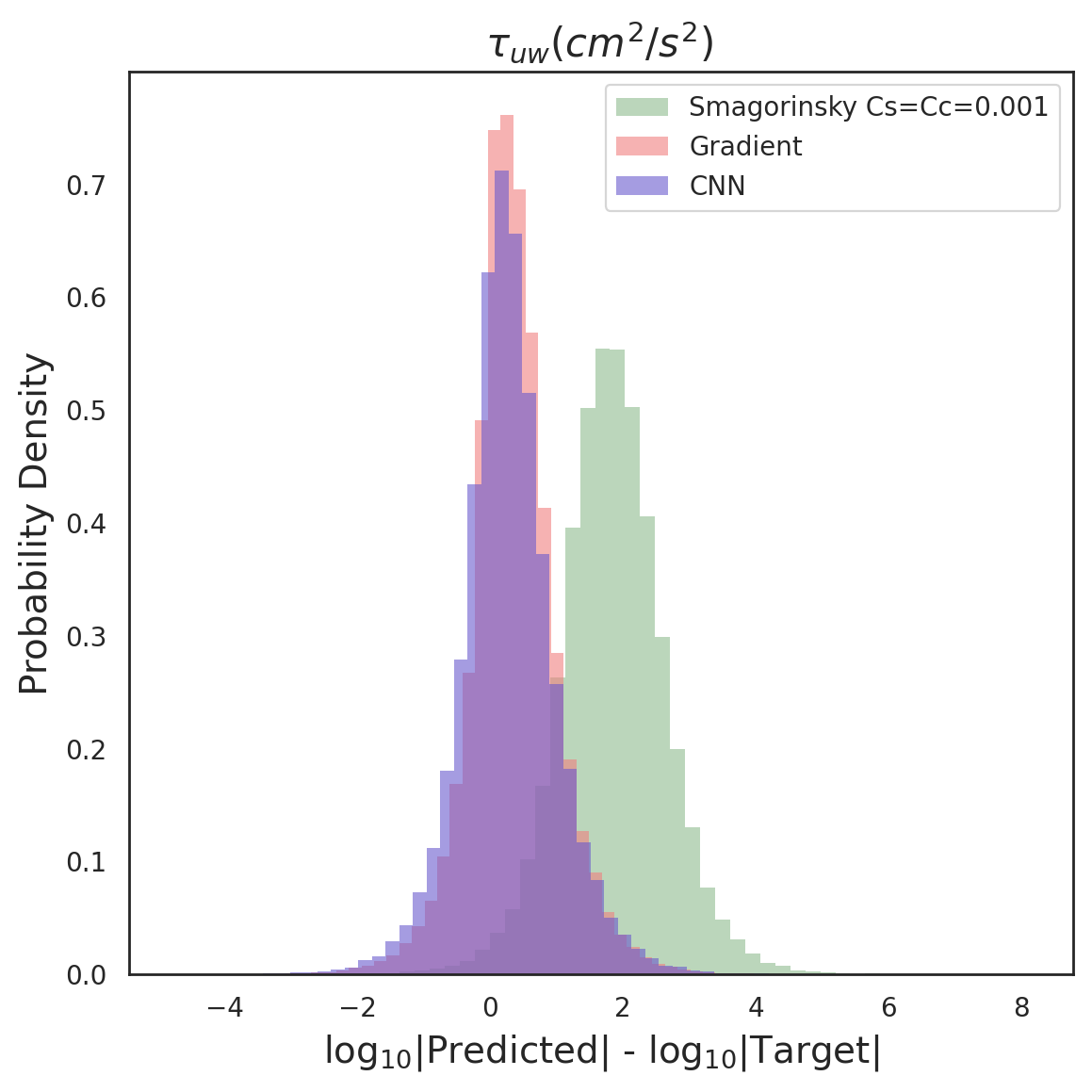}{0.48\textwidth}{\bfseries(b)}
}            
        \caption{ The Frequency plots of the deviations of $\tau_{vv}$ in Panel(a) and $\tau_{uw}$ in (b) between the predictions of the best CNN model (purple), Smagorinsky $C_S = C_C = 0.001$ (green), and Gradient models (red) and the target value.}
        \label{fig:ErrHist12_0001}
\end{figure}

We now examine the same error distributions for the Smagorinsky model with the reduced coefficient setting $C_S = C_C = 0.001$, shown in Figure~\ref{fig:ErrHist12_0001}. In contrast to the $C_S = C_C = 0.1$ case, the Smagorinsky model at this coefficient setting displays a markedly shifted error distribution, with its peak displaced significantly to the right of zero for both $\tau_{vv}$ and $\tau_{uw}$. This indicates a systematic overprediction of the stress tensor magnitudes by several orders of magnitude. The CNN error distribution, by contrast, remains tightly centered at zero. The discrepancy between the two Smagorinsky models additionally motivates the use of a data-driven surrogate that does not require coefficient specification.

\subsubsection{Scatterplots of Gradient and CNN Models}
\begin{figure}[tb]
    \centering
            \gridline{
  \fig{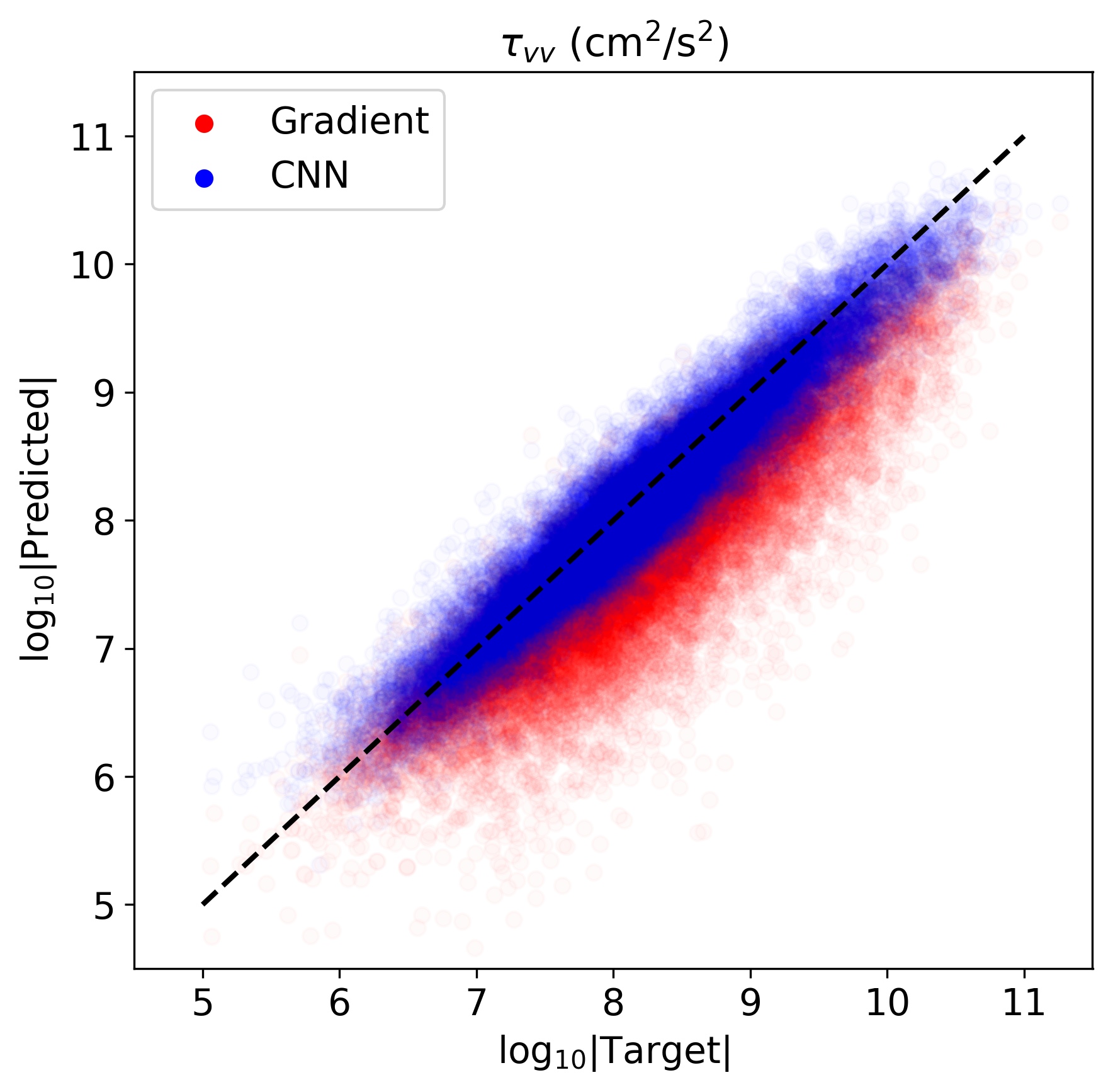}{0.48\textwidth}{\bfseries(a)}
  \fig{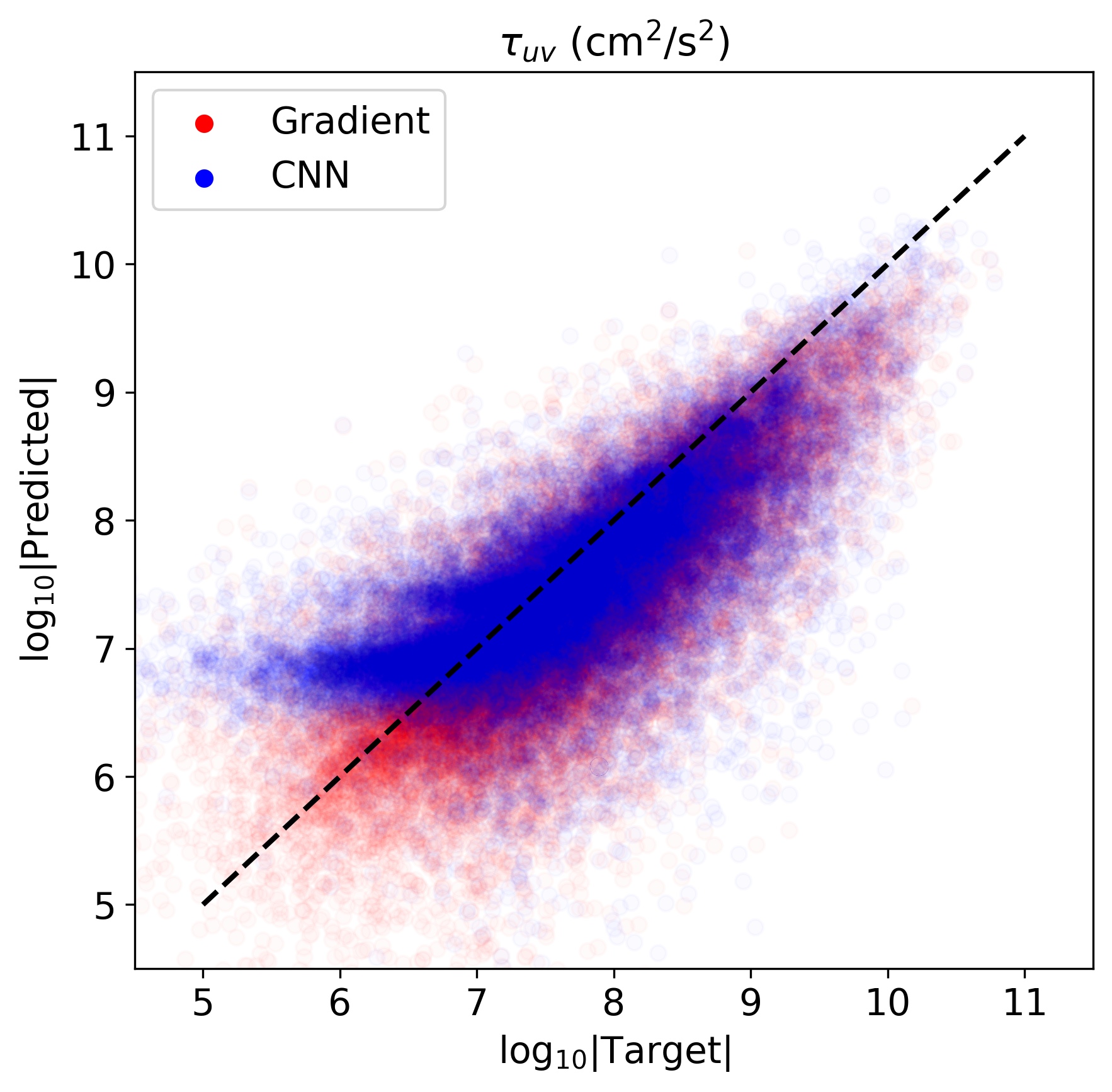}{0.48\textwidth}{\bfseries(b)}
} 
    \caption{The scatter plots of two components $\tau_{vv}$ in Panel(a) and $\tau_{uw}$ in (b) comparing the predictions of the best CNN model and the Gradient model. Blue points show the predictions of CNN, and red points show the predictions of the Gradient model.}
    \label{fig:GrVsCnn}
\end{figure}

Furthermore, we demonstrate the scatterplots in Figure~\ref{fig:GrVsCnn} comparing the predictions of the CNN model with the Gradient model, highlighting how each model performs across each sample in the testing dataset. The CNN model demonstrates a more consistent performance across the diagonal and off-diagonal Reynolds stress components, as evidenced by its tighter clustering of points around the diagonal dotted line representing a perfect prediction, particularly on $\tau_{vv}$ Figure ~\ref{fig:GrVsCnn}(a). Although CNN generally performs better than Gradient, there is a small accumulation of points where the predictions on $\tau_{uv}$ Figure ~\ref{fig:GrVsCnn}(b) deviate from the true values, particularly at the lower end of the target range. This suggests that the model may be underfitting in certain low-value regions, which could be due to insufficient representation of these values in the training data. Note that the plots of the components $\tau_{uu}$ and $\tau_{ww}$ were not displayed here because they demonstrate qualitatively similar behavior to  $\tau_{vv}$, while the plots of $\tau_{vw}$ and $\tau_{uw}$ plots have been omitted due to a similar behavior to $\tau_{uv}$. We should also note here that physically the sub-optimal performance of the model in the range of low values of Reynolds stress tensor components $\tau{}_{ij}$ is not as critical as its performance for the high values of stresses, since the feedback of the subgrid processes to the momentum and energy transport equations is, in general, proportional to the magnitude of the Reynolds stress tensor components. These results indicate that the CNN model is well-suited for predicting Reynolds stress components, offering more precise predictions than physics-based models. 

\subsection{ 3DCNN Evaluated on Low-Resolution Simulation Data}
To assess the practical applicability of the trained CNN model, we evaluate its performance when applied to velocity and density fields extracted from previously-introduced low-resolution ($\sim$50\,km, $C_S = C_C = 0.1$ and $C_S = C_C = 0.001$). This evaluation addresses the question of whether a model trained on high-resolution data generalizes to lower-resolution inputs, and whether the choice of Smagorinsky coefficient in the low-resolution simulation affects the CNN's predicted stress tensor distributions. We note here that this analysis is the evaluation of the learned closure, in which the surrogate is applied to resolved-scale fields from independent simulations rather than dynamically coupled into the governing equations of an active run. The full integration of the developed surrogate with the StellarBox simulations is the scope of our future work.

\subsubsection{Comparing CNN predictions on High vs Low Resolution}\label{section:lowres:predictions}
\begin{figure*}[tb]
        \centering
        \gridline{
  \fig{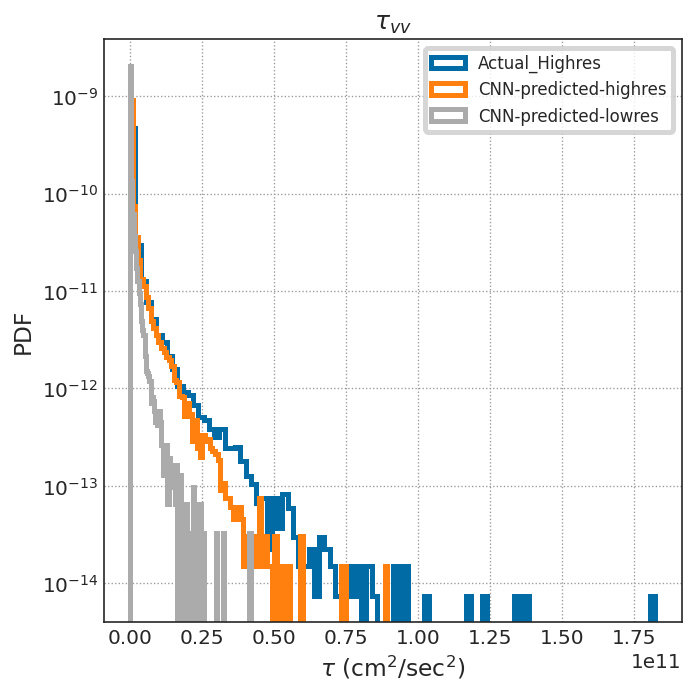}{0.48\textwidth}{\bfseries(a)}
  \fig{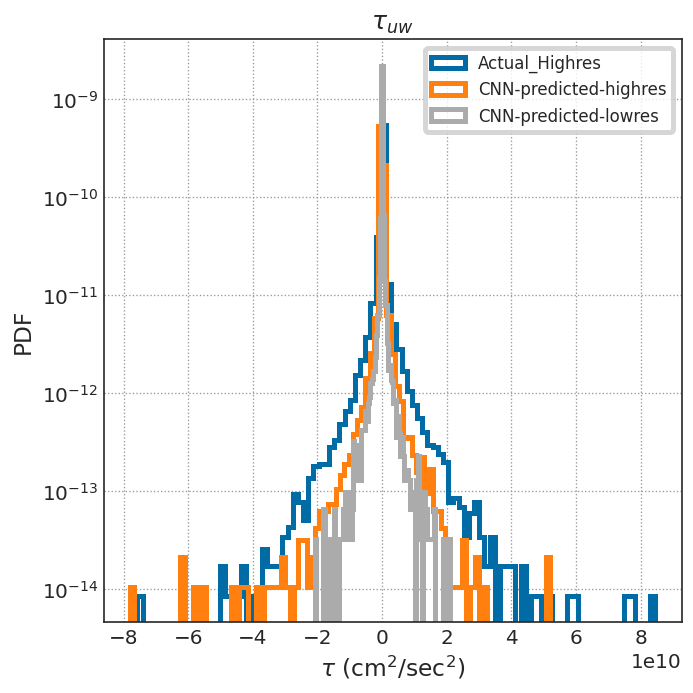}{0.48\textwidth}{\bfseries(b)}
}    
    \caption{Distributions of components $\tau_{vv}$ shown in Panel (a) and $\tau_{uw}$ in Panel (b). Blue curves show the original target distributions, orange curves show CNN predictions using filtered/coarse-grained inputs derived from the high-resolution simulations, and gray curves show CNN predictions using inputs from the genuine low-resolution simulations for the case $C_S = C_C = 0.001$.}
    \label{fig:PDFLowres0001}
\end{figure*}

\begin{figure*}[tb]
        \centering
        \gridline{
  \fig{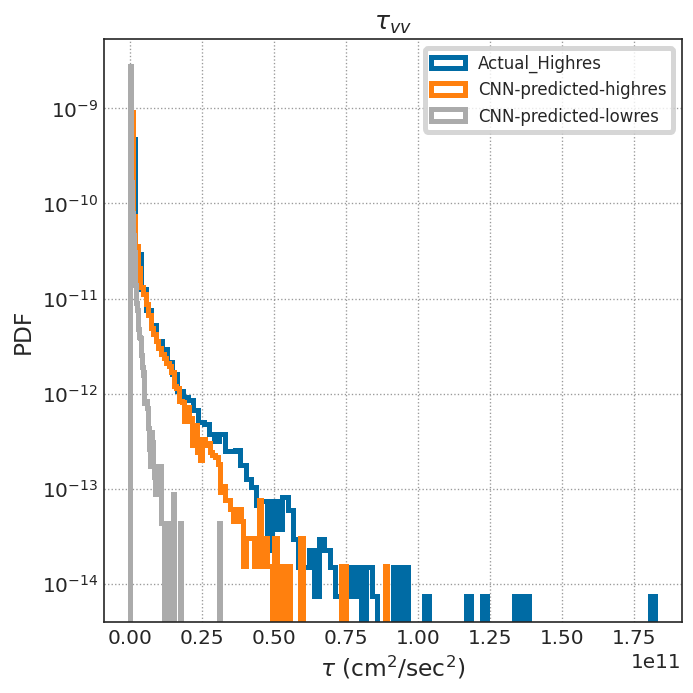}{0.48\textwidth}{\bfseries(a)}
  \fig{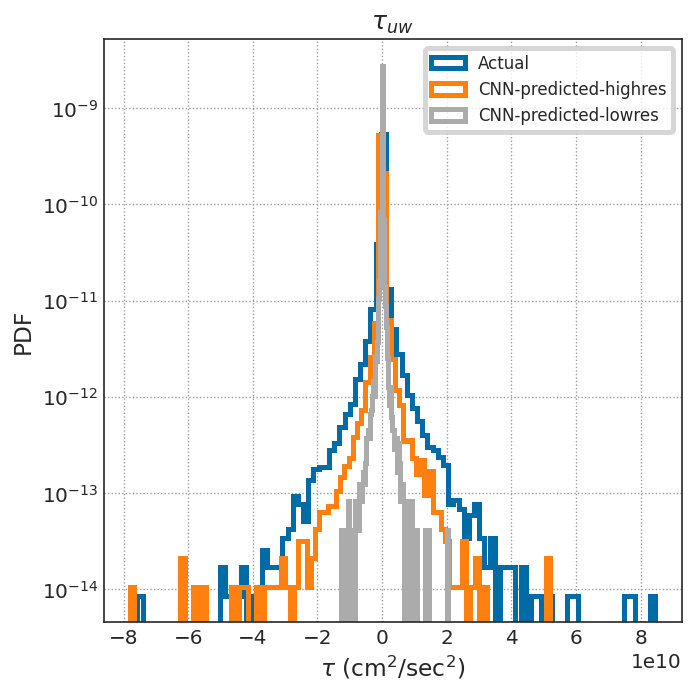}{0.48\textwidth}{\bfseries(b)}
}    
    \caption{Distributions of components $\tau_{vv}$ shown in Panel (a) and $\tau_{uw}$ in Panel (b). Blue curves show the original target distributions, orange curves show CNN predictions using filtered/coarse-grained inputs derived from the high-resolution simulations, and gray curves show CNN predictions using inputs from the genuine low-resolution simulations for the case $C_S = C_C = 0.1$.}
    \label{fig:PDFLowres01}
\end{figure*}

Figure~\ref{fig:PDFLowres0001} and Figure~\ref{fig:PDFLowres01} present the distributions of the CNN predicted Reynolds stress components $\tau_{vv}$ and $\tau_{uw}$ for the cases $C_S = C_C = 0.001$ and $C_S = C_C = 0.1$ low-resolution runs, respectively. Each figure shows three curves: the actual high-resolution target values (blue), the CNN predictions when fed high-resolution inputs (orange), and the CNN predictions when fed low-resolution inputs (gray). For clarity, the CNN predictions labeled as originating from the high-resolution simulations correspond to predictions obtained using filtered/coarse-grained inputs derived from the original high-resolution simulations, corresponding to an effective resolution of approximately $\sim$50\,km, rather than the original $\sim$12.5\,km, simulation. This is not a pointwise comparison; the low-resolution and high-resolution simulations do not correspond to the same spatial locations at the same time, but rather provide a comparison of the statistical distributions of the predicted values.

The most consistent observation across both coefficient settings and both stress components is that the CNN predictions on low-resolution input (gray) are noticeably narrower than both the actual target distribution (blue) and the high-resolution predictions (orange). The model captures the central bulk of the distribution well, correctly identifying the dominant near-zero region of the stress tensor, but underestimates the occurrence of larger-magnitude stress values in the tails. The high-resolution predictions (orange) align more closely with the actual target distribution (blue) across the full range, including the tails, as would be expected given that the model was trained on high-resolution data. The two low-resolution runs produce qualitatively similar prediction distributions, consistent with the modest differences in their input fields noted in Section~\ref{section:lowres:inputdist}. Subtle 
differences in the tail behavior between the two coefficient settings are visible and reflect the slightly broader velocity distributions of the $C_S = C_C = 0.001$ run propagating through the model.

\subsubsection{Discussion of Differences and Implications for Subgrid Modeling}\label{section:lowres:discussion}

The compression of the predicted stress distributions when the CNN is applied to low-resolution inputs, most likely, is a direct consequence of the differences in the input field distributions described in Section~\ref{section:lowres:inputdist}. The model was trained on high-resolution velocity fields that contain explicit small-scale fluctuations; when presented with low-resolution inputs in which those fluctuations have been averaged out, the model naturally predicts a narrower range of stress values. The observed distributional shift in the predictions is an expected consequence of the reduced variability in the low-resolution input fields. To further validate this point, it is essential to integrate the developed deep learning-based surrogate model into the StellarBox simulations, and then to compare the velocity and density distributions for the low-resolution simulations obtained using the surrogate model against those from the physics-based subgrid-scale models. We plan to approach it in our future work.

\subsection{Density's effect on Turbulence}

One can expect that the subgrid behavior could be different depending on the density environment (or the depth in the solar atmosphere). For example, the typical scale of the granulation is evident as granules of sizes of $\sim$1\, Mm on the surface and has approximately the same order-of-magnitude size of the convection cells with depth \citep{Stein1998granules}. We can, therefore, expect that at depths of $\sim$2\, Mm and below we can have a different structure of the flows, mostly controlled by larger convection scales. In this subsection, we will investigate the impact that the inclusion of the densities have on the Reynolds stress tensor modeling.

\begin{figure}[tb]
\centering
\includegraphics[width=\linewidth]{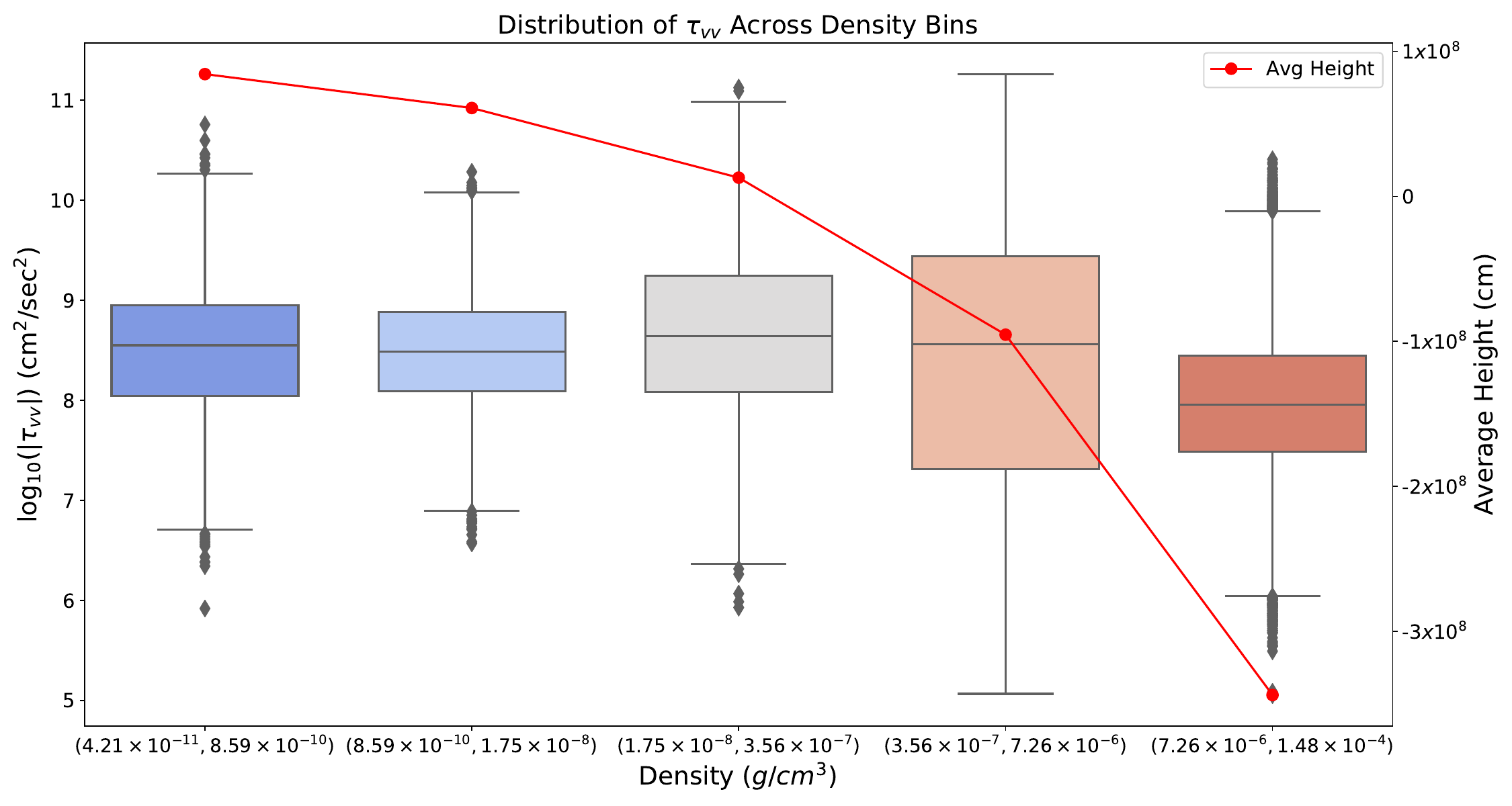}
\caption{The box-plots $\tau_{vv}$ across different density bins.}
\label{fig:TvvDen}
\end{figure}
\begin{figure}[tb]
\centering
\includegraphics[width=\linewidth]{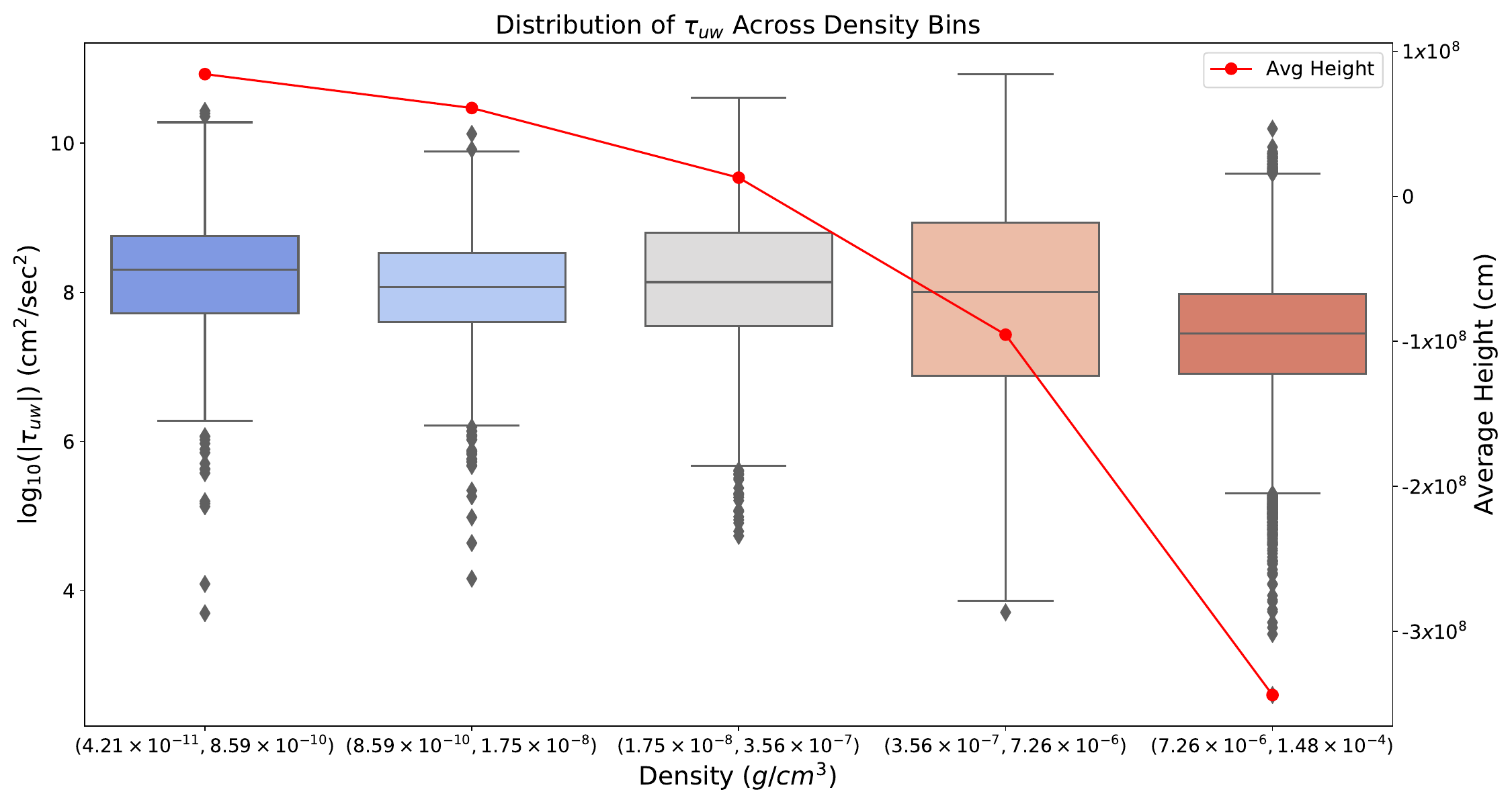}
\caption{The box-plots $\tau_{uw}$ across different density bins.}
\label{fig:TuwDen}
\end{figure}

The box-and-whisker plots of $\tau_{vv}$ and $\tau_{uw}$ components across different density bins are shown in Figures~\ref{fig:TvvDen}~and ~\ref{fig:TuwDen}. The plots reveal key trends in the behavior of turbulent stresses within the solar different solar densities and height environments by providing a statistical summary of the distribution of each variable across density bins. In such types of plots, the box represents the interquartile range (IQR), which spans the middle 50$\%$ of the data, while the horizontal line inside the box denotes the median value. The whiskers extend to the minimum and maximum values within 1.5 times the IQR, and outliers (data points lying beyond this range) are shown as individual markers. This format makes it easier to compare distributions, identify skewness, and detect extreme values. 

The variation in the median values and interquartile distances of $\tau_{vv}$ and $\tau_{uw}$ across the density bins in Figures~\ref{fig:TvvDen} and ~\ref{fig:TuwDen}, respectively, suggests a density-dependent relationship in the turbulent stresses. Lower-density bins generally exhibit smaller spreads (as indicated by narrower interquartile ranges). They most probably correspond to the layers in the solar atmosphere above the convection zone, where the convection motions are already suppressed. Conversely, higher-density bins tend to show increased variability, as demonstrated by wider interquartile ranges, suggesting more heterogeneous stress distributions in these regimes (besides the last bin, which likely corresponds to the depths of 2\, Mm and below, where the convective motions experience larger scale). Notably, the plots also highlight the presence of outliers across all density bins. These extreme values, particularly evident in both lower- and higher-density ranges, emphasize the inherent complexity of turbulence in solar conditions. These extreme values may result from sporadic and localized turbulent phenomena. Such features highlight the importance of robust modeling techniques that can capture the full range of variability while accounting for the heteroskedastic nature of the data.

\begin{figure}[htb]
    \centering
    \includegraphics[width=\linewidth]{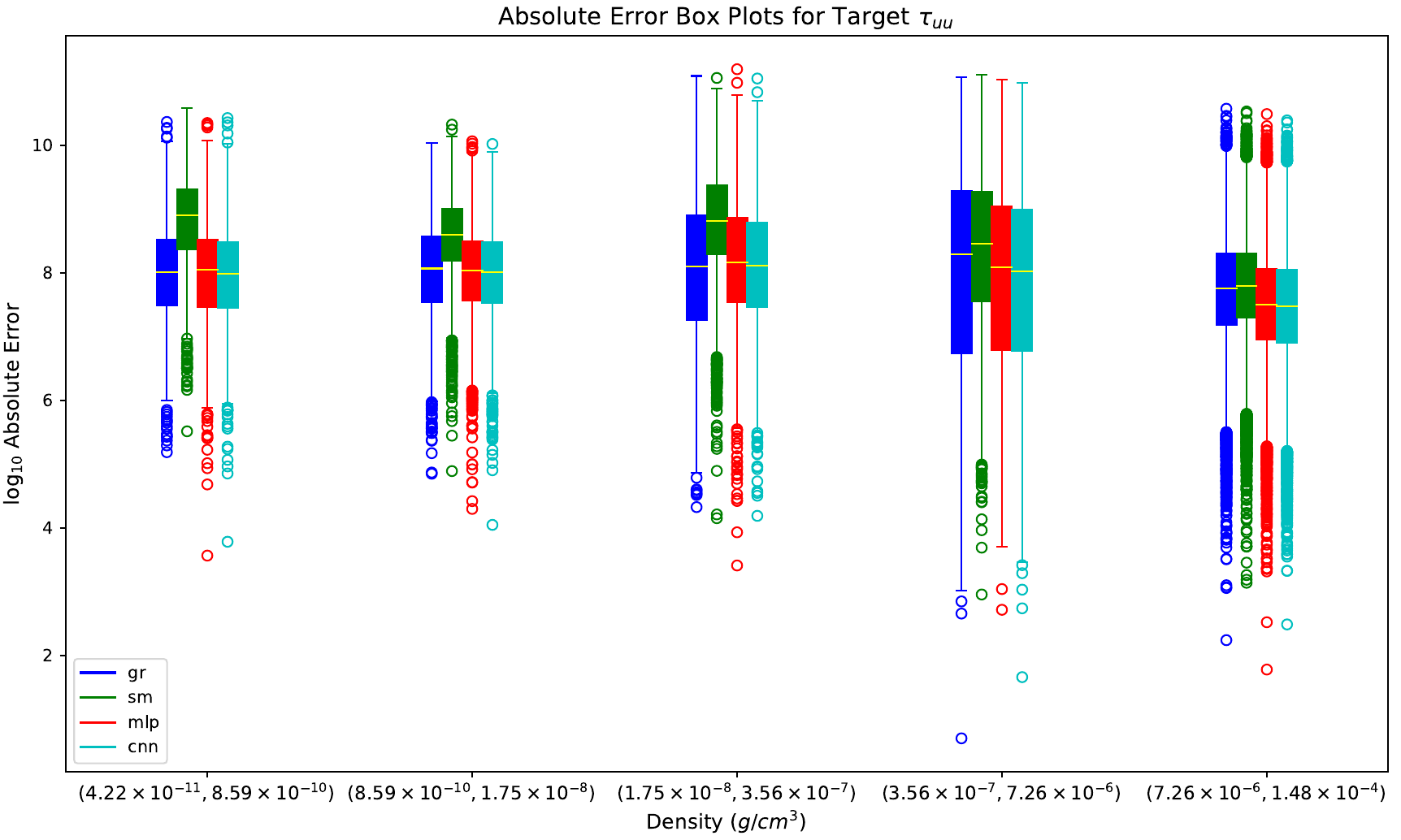}
    \caption{The error box-plots $\tau_{uu}$ across different density bins for all models(case Smagorinsky Cs=Cc=0.1).}
    \label{fig:errorBw1}
\end{figure}

The absolute error distributions of $\tau_{uu}$ across density bins, visualized in Figure~\ref{fig:errorBw1}, reveals important insights into model performance under varying density conditions. Among the models analyzed, the CNN consistently demonstrates superior performance, characterized by narrower interquartile ranges and fewer outliers across all bins, indicating its robustness and stability in capturing turbulence dynamics. The MLP performs comparably well but exhibits slightly higher variability and occasional outliers, particularly in lower-density regions. In contrast, the Gradient and Smagorinsky models display larger absolute errors and higher variability, especially in low-density bins, suggesting that these methods are less effective in handling the chaotic behavior associated with low-density turbulence. Notably, the absolute error variability decreases in higher-density bins for all models, reflecting improved predictability in these conditions. These results highlight the importance of robust models like CNNs for accurately predicting turbulence across a wide range of density conditions, with particular emphasis on addressing the challenges in low-density regions near the surface and in the solar atmosphere. These findings highlight the effectiveness of CNNs in modeling density-dependent turbulence dynamics and emphasize the challenges associated with low-density environments. The systematic variability observed across density bins reinforces the need for models that incorporate density-dependent features. 


\subsection{Data Transformation Effects}\label{section:transform_effects}
    
\begin{figure}[htb]
        \centering
                \gridline{
  \fig{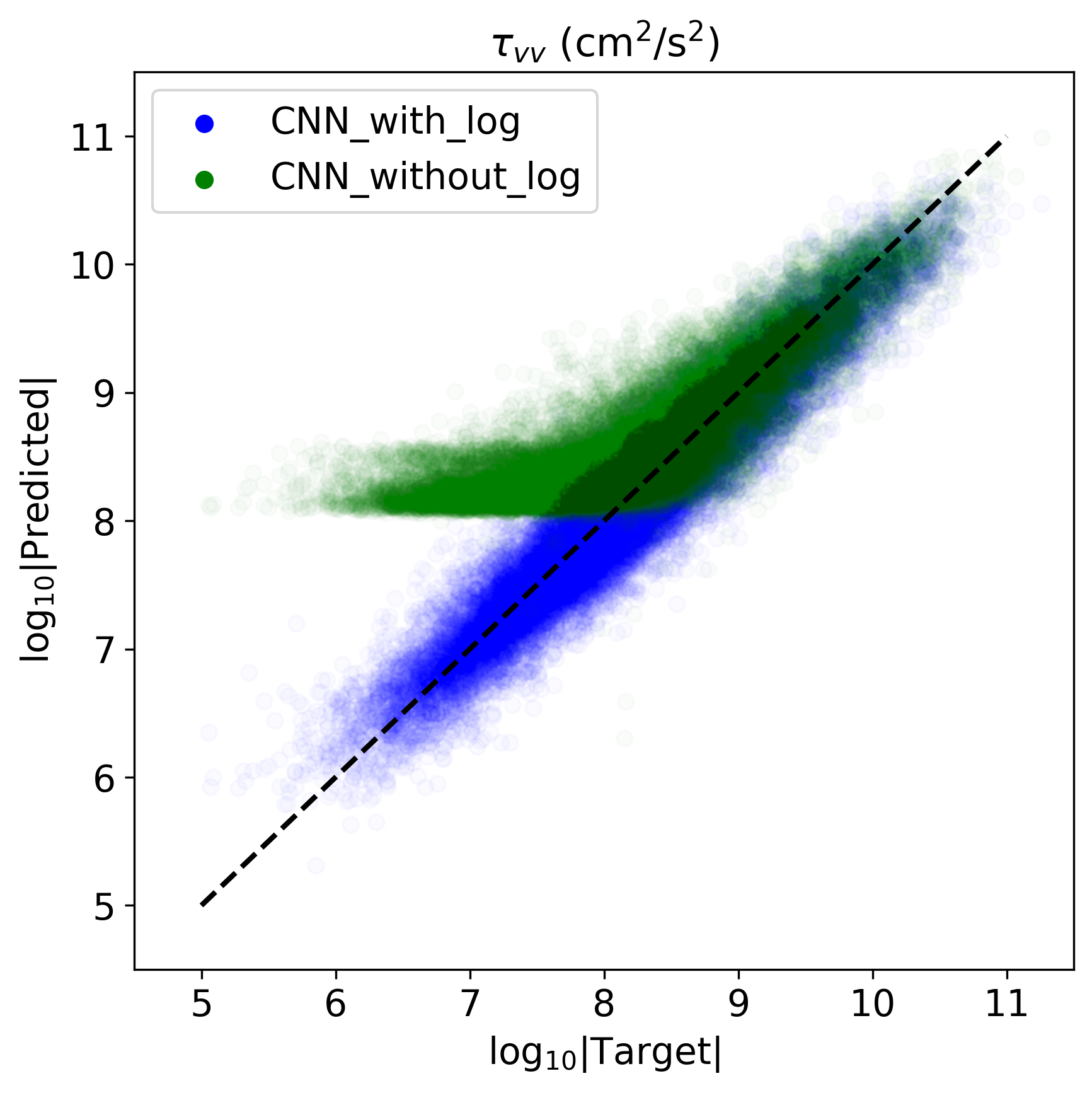}{0.48\textwidth}{\bfseries(a)}
  \fig{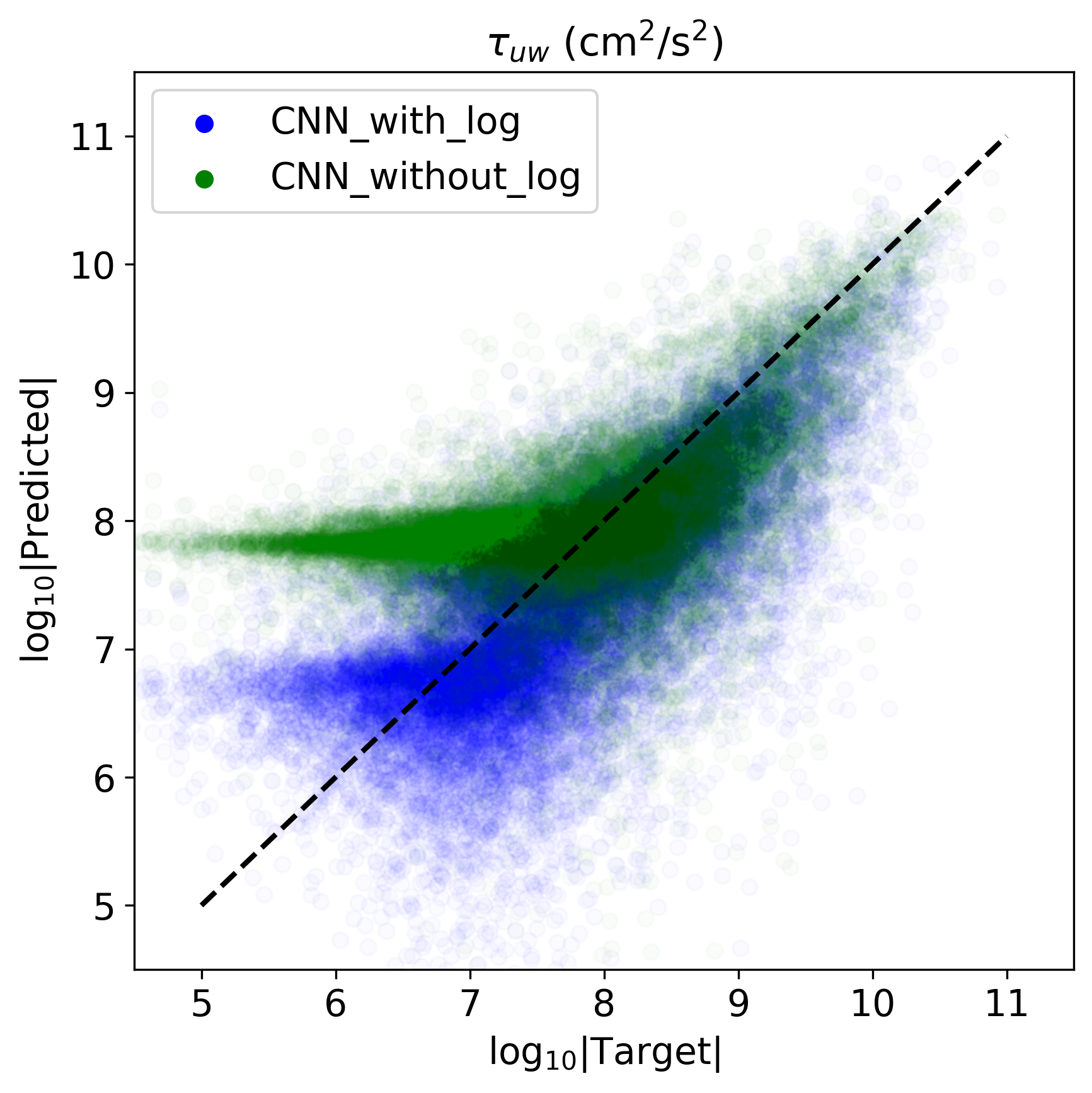}{0.48\textwidth}{\bfseries(b)}
} 
    
    \caption{ The scatter plots of two components $\tau_{vv}$ in Panel(a) and $\tau_{uw}$ in (b) comparing the predictions of the 3DCNN1 trained with data before and after normalization. Blue dots represent the predictions of CNN when trained with normalized data, and green represents the predictions of CNN when trained without a log transform.}
    \label{fig:NormVsOrg}
\end{figure}
We have investigated the performance of the selected CNN model trained using logarithmic transformation and scaled data, and using data that have not had a logarithmic transformation performed. In Figure \ref{fig:NormVsOrg}, we present two scatter plots of the predictions of these two CNN models. The CNN model trained on log-transformed data exhibits a much more robust and accurate prediction for both the $\tau_{vv}$ and $\tau_{uw}$ components, with significantly fewer outliers and better adherence to the $y=x$ line. This indicates that log transformation is an effective pre-processing step for stabilizing the training process and improving the prediction accuracy for Reynolds stress components that span several orders of magnitude. Without the log transformation, the scatter is more dispersed, and predictions deviate significantly from the diagonal.  The model struggles more with capturing the target values, particularly for small values, as the points cluster away from the diagonal. This is a direct consequence of the skewed distribution of the original data. The untransformed data is concentrated near zero, with long tails, making it difficult for the model to generalize.
\begin{deluxetable*}{ccccccccc}
\tablecaption{RMSE, RMSLE and $R^2$ scores comparing the CNN model when trained with log-scaled target features and without log-scaling. \label{table:CnnWOLog}}
\tablehead{
\colhead{\textbf{Model}} &
\colhead{\textbf{$\tau_{uu}$}} &
\colhead{\textbf{$\tau_{vv}$}} &
\colhead{\textbf{$\tau_{ww}$}} &
\colhead{\textbf{$\tau_{uv}$}} &
\colhead{\textbf{$\tau_{uw}$}} &
\colhead{\textbf{$\tau_{vw}$}} &
\colhead{\textbf{RMSLE}} &
\colhead{\textbf{$R^2$}}
}
\startdata
CNN\_with\_Log     & $2.36\times10^{9}$ & $2.50\times10^{9}$ & $3.13\times10^{9}$ & $1.42\times10^{9}$ & $1.67\times10^{9}$ & $1.70\times10^{9}$ & 1.31 & 0.54 \rowsp\\
CNN\_without\_Log  & $2.28\times10^{9}$ & $2.27\times10^{9}$ & $2.81\times10^{9}$ & $1.31\times10^{9}$ & $1.51\times10^{9}$ & $1.51\times10^{9}$ & 1.80 & 0.62 \rowsp\\ \rowspb
\enddata
\end{deluxetable*}



As follows from Table~\ref{table:CnnWOLog}, the CNN trained without performing log-transformation on the targets demonstrated lower RMSE values on all the 6 components and a higher R$^2$ score. However, Figure~\ref{fig:NormVsOrg} demonstrates that the model trained without performing a log transformation on the data predicts well only on the higher values, and fails in the lower ranges. This could be because all the targets are heavily centered near zero. Therefore, when a StandardScaler is applied to this data, the distribution remains the same, with most data points cluttered near zero. On deeper analysis of the Standard scaled targets, it was observed that $75\%$ of the data is lower than $ 10^{-1}$ (in the transformed scale), and around $25\%$ of the data has a value higher than 1, which constitutes the outliers. The average RMSLE is also lower on the model trained with log-transformed data. This is because when untransformed data is fed to the model for training, the model underfits on values close to zero and prioritizes the higher values (outliers). This model only performs well on the outliers and underperforms on most of the data. This pattern is consistent with other studies \citep[e.g.,][]{OLIVIER2008333,bellego2022, Cleophas2016} showing that log transformations, particularly natural log, are highly effective when data distributions are heavily skewed, improving linearity between features and stabilizing variance.
\subsection{Preliminary Cluster Analysis as Motivation for Further Work}
To better understand the variability and predictability of all Reynolds stress tensor components ($\tau_{uu}$, $\tau_{uv}$, $\tau_{uw}$, $\tau_{vv}$, $\tau_{vw}$, $\tau_{ww}$), a K-Means clustering analysis was applied on the test dataset. Using the 50-km averaged velocity components ($u$, $v$, $w$) from a 3$\times$3$\times$3 domain data cubes and the values of the density in the central cube, the data was clustered into 5 groups (labeled Cluster~0 to Cluster~4), to identify patterns or regions in the data with similar turbulent characteristics. The clustering was done in the Euclidean space on the features normalized to the mean and scaled with the standard deviation, and has been tested only on the first considered data cube of the simulations. The clustering process grouped observations based on feature similarities, effectively segmenting the data into clusters with distinct behaviors. The Mean Squared Error (MSE) and $R^{2}$ scores were subsequently evaluated within each cluster to assess the model's predictive performance for the Reynolds stress tensor components as seen in Table~\ref{table:Cl_MSE_R2}.

\begin{deluxetable}{cccc}
\tablecaption{Cluster-wise sample counts, average MSE, and $R^{2}$ scores. \label{table:Cl_MSE_R2}}
\tablehead{
\colhead{\textbf{Cluster}} & 
\colhead{\textbf{Number of Points}} & 
\colhead{\textbf{Average MSE}} & 
\colhead{\textbf{Average $R^{2}$}}
}
\startdata
0 & 11,134 & $4.72 \times 10^{18}$ & 0.576 \rowsp\\
1 & 39,752 & $3.28 \times 10^{17}$ & 0.563 \rowsp \\
2 & 6,598  & $3.66 \times 10^{19}$ & 0.509 \rowsp\\
3 & 8,912  & $3.60 \times 10^{18}$ & 0.595 \rowsp\\
4 & 7,188  & $8.80 \times 10^{18}$ & 0.547 \rowsp\\ \rowspb
\enddata
\end{deluxetable}

The clustering analysis revealed distinct regimes in the dataset, each associated with characteristic density, velocity, and prediction accuracy patterns. Table~\ref{table:cluster_stats} summarizes the mean and standard deviation of the density and velocity of the five identified clusters. Cluster 1, the largest cluster (39,752 points), corresponding to the most dense and slowest flowing plasma(highest mean density and lowest mean velocity), exhibits the lowest mean squared error (MSE $\sim 3.3 \times 10^{17}$) with an average $R^{2}$ score of 0.56. This indicates that stable high-density regions (that correspond to the deeper solar interior) are comparatively easier to predict, and the model captured the variance and scales more effectively.
\begin{deluxetable*}{ccccc}
\tablecaption{Cluster-wise mean and standard deviation of density (g/cm$^{3}$) and vertical velocity (cm/s). \label{table:cluster_stats}}
\tablehead{
\colhead{\textbf{Cluster}} &
\colhead{\textbf{Mean Density}} &
\colhead{\textbf{Standard Deviation Density}} &
\colhead{\textbf{Mean Velocity}} &
\colhead{\textbf{Standard Deviation Velocity}}
}
\startdata
0 & $7.28 \times 10^{-6}$ & $1.30 \times 10^{-5}$ & $9.39 \times 10^{4}$ & $7.27 \times 10^{4}$ \rowsp\\
1 & $5.73 \times 10^{-5}$ & $4.20 \times 10^{-5}$ & $4.64 \times 10^{4}$ & $3.87 \times 10^{4}$ \rowsp\\
2 & $6.45 \times 10^{-6}$ & $1.50 \times 10^{-5}$ & $2.79 \times 10^{5}$ & $1.49 \times 10^{5}$ \rowsp\\
3 & $2.04 \times 10^{-6}$ & $3.00 \times 10^{-6}$ & $1.05 \times 10^{5}$ & $8.10 \times 10^{4}$ \rowsp\\
4 & $8.75 \times 10^{-7}$ & $3.00 \times 10^{-6}$ & $1.06 \times 10^{5}$ & $8.53 \times 10^{4}$ \rowsp\\ \rowspb
\enddata
\end{deluxetable*}


It is important to analyze other clusters as well. Cluster 0 (11,134 points) and Cluster 4 (7,188 points) show a moderate variability in density and velocity values, with average $R^{2}$ scores of 0.55-0.57 but elevated error magnitudes ($\sim 10^{18}$–$10^{19}$). These clusters suggest that, while the model can capture relative variance, it struggles with scaling predictions in transitional layers of the atmosphere. Cluster 3 (8,912 points), which represents very low-density and moderate-velocity conditions, achieves the highest $R^{2}$ score 0.59 despite a non-negligible MSE ($\sim 3.6 \times 10^{18}$). This suggests that while absolute errors persist, the model effectively learns relative variability in these regions. In contrast, Cluster 2 (6,598 points), characterized by low mean density and extremely high mean velocity, displays the highest prediction errors (MSE $\sim 3.7 \times 10^{19}$), the lowest $R^{2}$ (0.51), and a high velocity standard deviation, representing a highly dynamic and turbulent plasma. This preliminary analysis provides motivation for adjustments of the models training process depending on the cluster the points belong to, with the potential to further improve the models.

\section{Conclusion}\label{section:conclusion}

In this work, we developed an MLP Regressor and two variants of a 3D Convolutional Neural Network (CNN) namely 3DCNN1 and 3DCNN2 to predict the subgrid Reynolds Stress Tensor components $\tau_{ij}$ in high-resolution 3D radiative hydrodynamic simulations of the upper convection zone (up to $\sim$5.4\, Mm beneath the surface) and the lower atmosphere (up to $\sim$1\,Mm above the surface) of the Sun. The simulation was initially performed with the high spatial resolution of $\sim$12.5\, km. The input features comprised the 3D velocity components averaged over the spatial resolution of 50\, km along with the mean density, incorporated differently in each CNN version. 
Both CNNs were parameter-tuned extensively and compared against each other, as well as an MLP Regressor and physics-based Gradient and Smagorinsky baselines. A comprehensive performance analysis involved evaluating predictions via log-transformed scatter plots, error histograms, and probability density function (PDF) comparisons. The MLP demonstrated improvements over the Gradient baseline on the diagonal components, though it lagged in off-diagonal predictions. Among all models, 3DCNN1 (Design 1.1) consistently achieved lower MSEs and higher R$^2$ scores, slightly outperforming 3DCNN2 (Design 2.1), which was likely hampered by the increased data dimensionality. These CNNs have also shown lower RMSE scores on all diagonal and off-diagonal components of $\tau$ compared to MLP and the baseline. Critically, log-transforming the heavily skewed target data proved essential for capturing both small- and large-magnitude stresses, as models trained on only standard-scaled data tended to underfit near-zero values, prioritizing outliers instead.
Additional analyses provided deeper insights into model behavior: 
\begin{itemize}[leftmargin=*] 
\item Smagorinsky Coefficient Sensitivity: Evaluating the Smagorinsky model under two coefficient settings ($C_S = C_C = 0.1$ and $C_S = C_C = 0.001$) revealed that the physics-based baseline is highly sensitive to this parameter. The CNN surrogate does not require a selection of coefficients, and therefore has an advantage in this respect.

\item Density Effects: Binning predictions by density revealed that low-density regions, which likely correspond to the upper convection zone and lower atmosphere, pose stronger modeling challenges due to their larger variability and extreme outliers. All models showed wider error distributions in these bins, though 3DCNN1 maintained a comparatively narrower spread. 

\item Low-Resolution Generalization: When applied to low-resolution simulation inputs, the CNN predictions show a narrowing of the predicted stress distributions relative to the high-resolution case, which is, most likely, a consequence of the smoother velocity fields at coarser resolution. Despite this, the CNN recovers the distributional shape of the stresses more reliably than physics-based models considered, supporting its potential utility as a subgrid surrogate in practical low-resolution simulation workflows.

\item Cluster Analysis: Segmentation based on K-means of the test dataset (based on the averaged velocity fields and density) into five clusters illustrated that certain clusters (e.g., Cluster 0,1) consistently yielded low MSEs across stress components, whereas others (e.g., Cluster 2) were persistently difficult to predict. This highlights the heterogeneous nature of solar turbulent flow fields and underscores the value of localized performance evaluations. 
\end{itemize}

Importantly, this work demonstrated the potential of deep learning techniques for the development of the subgrid turbulent transport models of the high-resolution simulations of the Sun. In particular, we observe that the best-performing deep learning models (summarized in Table~\ref{table:results}) outperform the baseline Gradient and Smagorinsky models in estimating all Reynolds Stress Tensor components. The performance is especially notable to the diagonal components of the stress tensor. In particular, the RMSE scores have decreased for about $\sim$30\% on average for the diagonal components with respect to $\sim$8\% decrease for the off-diagonal components if comparing the 3DCNN1 model with the gradient physics-based model. The improvement relative to the Smagorinsky model is even larger for the $C_S = C_C = 0.001$ setting, where the physics-based baseline deviates substantially from the target values across all components (see Table~\ref{table:results} for more details). Figures~\ref{fig:ErrHist12}~and~\ref{fig:errorBw1} show that the estimates based on the CNN model are less biased and less spread. We have also evidently illustrated that not all deep learning architectures work similarly for the subgrid stress tensor estimation and that the 3D CNN structures show more promise with respect to the `conventional' MLPs in this area.

While the CNN model demonstrates strong performance across various Reynolds stress components, there are potential areas for improvement. One such area could include a loss weighting schema based on clustering to address data imbalance among regions of interest, ensuring that underrepresented data is given appropriate emphasis in model training. Secondly, generalization techniques like L2 regularization (weight decay) and dropout can help reduce overfitting and enhance the model's ability to capture both high and low-probability regions, and incorporating data augmentation may further enhance the model’s ability to generalize to more complex flow regimes. Finally, incorporating ensemble learning techniques by combining the CNN with other neural architectures, such as recurrent neural networks (RNNs) or transformers, could provide additional performance gains in the prediction of turbulent transport phenomena.

\section{Data Availability}
The simulation dataset used in this study was generated using the StellarBox code. The derived ML training dataset will be made available upon request.

\section{software}
PyTorch \citep{paszke2019pytorchimperativestylehighperformance},
NumPy \citep{harris2020numpy},
SciPy \citep{2020SciPy-NMeth},
scikit-learn \citep{scikit-learn},
Matplotlib \citep{hunter2007matplotlib},
Pandas \citep{reback2020pandas},
StellarBox \citep{Wray2015_stellarbox, Wray2020}

\section{Acknowledgements}

This work was partially supported by NASA's Diversify, Realize, Integrate, Venture, Educate (DRIVE) Science Center grant \#80NSSC22M0162 of the Consequences of Fields and Flows in the Interior and Exterior of the Sun (COFFIES) center. VMS also thanks the NSF FDSS grant \#1936361.


\bibliographystyle{aasjournal}
\bibliography{refer} 
\end{document}